\documentclass{aa}  
\usepackage[utf8]{inputenc}
\usepackage{graphicx}
\usepackage{caption}
\usepackage{subcaption}
\usepackage{float}
\usepackage{txfonts}
\usepackage{natbib}
\usepackage{xcolor}
\usepackage{bm}
\usepackage{hyperref}
\usepackage{multirow}
\hypersetup{
    colorlinks = true,
    citecolor = {blue},
    linkcolor = {blue},
    urlcolor = {blue}
}
\newcommand{\ii}{ {\rm i} }
\newcommand{\id}{ {\rm d} }
\newcommand{\ephi}{{\hat{\bm \phi}}}
\newcommand{\etheta}{{\hat{\bm \theta}}}
\newcommand{\er}{{\hat{\bm r}}}
\newcommand{\ez}{{\hat{\bm z}}}
\newcommand{\br}{{\bm r}}

\newcommand{\Omegaref}{\Omega_{\rm ref}}

\newcommand{\bu}{ \boldsymbol{u} }
\newcommand{\bU}{ \boldsymbol{U} }

\newcommand{\vort}{\zeta}

\begin{document} 
\title{Viscous inertial modes   on a differentially rotating sphere:\\ Comparison with solar observations}

\titlerunning{Viscous inertial modes on a differentially rotating sphere}

\authorrunning{Fournier et al.}
\author{
Damien Fournier\inst{1} \and Laurent Gizon\inst{1,2,3} \and
Laura Hyest\inst{1,4}
}

\institute{
Max-Planck-Institut f\"ur Sonnensystemforschung, Justus-von-Liebig-Weg 3, 37077 G\"ottingen, Germany 
    \and
Institut f\"ur Astrophysik, Georg-August-Universit\"at G\"ottingen,  Friedrich-Hund-Platz 1, 37077 G\"ottingen, Germany
    \and
Center for Space Science, NYUAD Institute, New York University Abu Dhabi, Abu Dhabi, UAE
\and
Institut Sup\'erieur
de l'A\'eronautique et de l'Espace -- SUPAERO,
10 avenue \'Edouard Belin,
31055 Toulouse, France
}

\date{Received \today ;  Accepted <date>}

 
  \abstract
   {In a previous paper we studied the effect of latitudinal rotation on solar equatorial Rossby modes in the $\beta$-plane approximation. Since then, a rich spectrum of inertial modes has been observed on the Sun, which is not limited to the equatorial Rossby modes and includes high-latitude modes.}
   {Here we extend the computation of toroidal modes in 2D to spherical geometry, using realistic solar differential rotation and including viscous damping. The aim is to compare the computed mode spectra with the observations and  to study mode stability.} 
   {At fixed radius, we solve the eigenvalue problem numerically using a spherical harmonics decomposition of the velocity stream function.}
   {Due to the presence of viscous critical layers, the spectrum consists of four different families: Rossby modes, high-latitude modes, critical-latitude  modes, and strongly damped modes. For each longitudinal wavenumber $m\leq3$, up to three Rossby-like modes are present on the sphere, in contrast to the equatorial  $\beta$ plane  where only the equatorial Rossby mode is present. The least damped modes in the model have eigenfrequencies and eigenfunctions that resemble the observed modes; the comparison improves when the radius is taken in the lower half of the convection zone. 
   For radii above $0.75R_\odot$ and Ekman numbers $E<10^{-4}$, at least one mode is unstable. For either $m=1$ or $m=2$, up to two Rossby modes  (one symmetric, one antisymmetric) are unstable when the radial dependence of the Ekman number  follows a quenched diffusivity model ($E\approx 2\times 10^{-5}$ at the base of the convection zone).
   For $m=3$, up to two Rossby modes can be unstable, including the equatorial Rossby mode.
  }
   {
   Although the 2D model discussed here is highly simplified, the  spectrum of toroidal modes appears to include many of the observed solar inertial modes. The self-excited modes in the model have frequencies close to those of the observed modes with the largest amplitudes. 
   }


\maketitle


\section{Introduction}

\citet{Gizon2021} reported the observation and identification of a large set of global modes of solar oscillations in the inertial frequency range. These include the equatorial Rossby modes \citep{Loeptien18},  high-latitude modes  \citep[the $m=1$ symmetric mode manifests itself as a spiral structure, see][]{Hathaway2013,Bogart2015}, and critical-latitude modes.
The observed modes were identified by comparison with (a) inertial modes
computed for an axisymmetric model of the convection zone (using a 2D solver)
and (b)  purely toroidal modes computed on the solar surface (using a 1D solver).
In this paper, we provide additional results based on the 1D solver.
Additional results based on the 2D solver are provided in a companion paper \citep{Bekki2022}.

Under a simplified setup in the equatorial $\beta$ plane, \citet{Gizon2020} discussed the importance of latitudinal differential rotation in this problem. The eigenvalue problem for  purely toroidal modes is singular in the absence of viscosity. In the presence of eddy viscosity, the eigenmodes can be grouped into several families of modes: the critical-latitude modes, the high-latitude modes, the strongly damped modes, and the equatorial Rossby modes. These modes arise from the existence of viscous critical layers where the wave speed is equal to the zonal shear flow.  In particular, the equatorial Rossby modes (the R modes) are trapped below the critical layer.
In the present paper, we wish to extend the work of \citet{Gizon2020} to the study of toroidal modes on the sphere (i.e. we drop the equatorial  $\beta$-plane approximation) in the presence of realistic latitudinal differential rotation and eddy viscosity. The intention is to include critical and high-latitude modes in the discussion: can these modes also be captured by simple physics on the sphere?

Instabilities and waves on a differentially rotating sphere have been studied before. The basic equation for the oscillations (Sect.~\ref{sect:linearModes}) was established (in an inertial frame) by \citet{Watson1981} and later extended to more realistic solar rotation profiles by \citet{Dziembowski87c} and \citet{Charbonneau1999}. All these studies showed the possibility of unstable modes when the differential rotation is strong enough, which is the case for the Sun in the upper convection zone. Unstable modes also appear in the overshooting part of the tachocline when considering a shallow-water model allowing for radial motions instead of purely toroidal modes \citep{Dikpati2001}. 

Here, we restrict ourselves to the 2D approximation which is valid for strongly stratified media when the rotation rate is much smaller than the buoyancy frequency  \citep{Watson1981}. For the Sun, it is valid in the radiative zone, the atmosphere, and the outer part of the convective envelope \citep{Dziembowski87c}. Moreover, 
the 2D approximation is a valuable approximation to the 3D problem for the modes that vary slowly with depth
\citep{Kitchatinov2009} and are away from the axis of symmetry \citep{Rieutord02}. 

In this work we study only the hydrodynamical modes and  do not consider the influence of a magnetic field, even if it was shown to be a significant ingredient  \citep{Gilman2000,Gilman2002,Gilman2007}. The main difference with previous works is the introduction of a viscous term which is important to understand the shape of the eigenfunctions and the lifetime of the subcritical modes.
Also we do not restrict latitudinal differential rotation to a two- or four-term profile, but consider  the  solar rotation profile as measured by helioseismology at the solar surface (Sect.~\ref{sect:surfaceDiffRot}) and in the interior (Sect.~\ref{sect:depth}). We study the spectrum and the eigenfunctions of the normal modes of the system.
Depending on the parameters of the problem, we find that some modes may be self-excited (Sect.~\ref{sect:depth}).
The conclusion, Sect.~\ref{sect:conclusion}, includes a short discussion on the stability of differential rotation for distant stars.  

\section{Linear toroidal modes on a sphere} \label{sect:linearModes}
\subsection{Equation for the velocity stream function}

We study the propagation of purely toroidal modes on a sphere of radius $r$ under the influence of latitudinal differential rotation. In an inertial frame, the rotation rate  $\Omega(\theta)$ depends on  the co-latitude $\theta$ measured from the rotation axis $\ez$.  We work in a frame rotating at the reference angular velocity $\Omega_{\rm ref}$ (chosen to be the Carrington rate later in the paper). In the rotating frame, the Navier-Stokes equation   is
\begin{equation}
    \frac{\partial\bu}{\partial t}  + \bu \cdot \nabla  \bu +  2 \Omegaref \ez \times \bu
    = {\nabla \Pi} + \nu  \Delta \bu,  \label{eq:Momentum}
\end{equation}
where $\ez$ is the unit vector along the rotation axis, $\bu$ the horizontal  velocity, and  $\nu$ the eddy viscosity. 
 For the sake of simplicity, wave damping by  turbulence  is modelled by a horizontal Laplacian $\Delta$.
The  force on the  right-hand side is assumed to derive  from a potential $\Pi$. 
In the rotating frame, we decompose the velocity into the mean axisymmetric flow $\bU$ and the wave velocity $\bu'$,
\begin{eqnarray}
\bu(\theta,\phi,t) = \bU(\theta) + \bu'(\theta,\phi,t),
\end{eqnarray}
where  $\phi$ is the longitude (increases in the prograde direction). 
For example,  $\bU=(\Omega - \Omegaref) \ez \times \br$ is the flow associated with latitudinal differential rotation. 
To first order in the wave amplitude, we have
\begin{equation}
    \frac{\partial\bu' }{\partial t} 
    + (\bU \cdot \nabla ) \bu' 
    + (\bu' \cdot \nabla )\bU
    +  2 \Omegaref \ez \times \bu' = -\nabla \Pi' + \nu  \Delta \bu'.  \label{eq:Momentum1}
\end{equation}
The two horizontal components of this equation are
\begin{align}
 & D_t  u'_\theta - 2 \Omega \cos\theta\ u'_\phi  =  -\frac{1}{ r} \frac{\partial \Pi'}{\partial\theta}
 + \nu  \Delta u'_\theta, \label{eq:momentumTh} \\
 & D_t  u'_\phi  + \frac{1}{\sin\theta} 
 \frac{\id}{\id\theta} \left( \Omega \sin^2\theta \right) u'_\theta=
 - \frac{1}{r\sin \theta} \frac{\partial\Pi'}{\partial \phi}
 + \nu  \Delta u'_\phi,
 \label{eq:momentumPhi}
\end{align}
where 
\begin{equation}
D_t = \frac{\partial}{\partial t} + (\Omega-\Omegaref) \frac{\partial}{\partial \phi}
\end{equation} is the material derivative in the rotating frame.
For purely toroidal modes, we can introduce the stream function $\Psi(\theta,\phi,t)$ such that
\begin{equation}
    \bu' = \nabla \times \bigl[ \Psi(\theta,\phi,t)\ \er \bigr] = \frac{1}{r \sin\theta} \frac{\partial\Psi}{\partial \phi}\ 
    \etheta - \frac{1}{r} \frac{\partial\Psi}{\partial \theta}\ \ephi.
\label{eq: u'}
\end{equation}
Combining Eq.~\eqref{eq:momentumTh} and Eq.~\eqref{eq:momentumPhi}, we obtain the equation of \citet{Watson1981} modified on the right-hand side to include  viscosity:
\begin{equation}
D_t \Delta \Psi -  \frac{\zeta \ \Omegaref}{r^2} \frac{\partial\Psi}{\partial \phi}
= \nu \Delta^2\Psi, \label{eq:Watson}
\end{equation}
where 
\begin{equation}
    \zeta(\theta) =   \frac{1}{\Omegaref \sin\theta} \frac{\id}{\id\theta}  \left( \frac{1}{\sin\theta} \frac{\id}{\id\theta}  (\Omega \sin^2\theta) \right) .
    \label{eq:zeta}
\end{equation}

\subsection{Modal decomposition}

We look for wave solutions of the form
\begin{equation}
    \Psi(\theta,\phi,t) = \textrm{Re} \bigl[ \psi(\theta) \textrm{e}^{\ii(m\phi - \omega t)} \bigr],
\label{Eq:Psi}
\end{equation}
where $m$ is the longitudinal wavenumber and $\omega$ is the
(complex) angular frequency measured in the rotating reference  frame. 
 Introducing the Ekman number 
\begin{equation}
E=\frac{\nu}{r^2\Omegaref},  \label{eq:Ekman}
\end{equation}
the function $\psi$ satisfies 
\begin{equation}
(m \delta  - \omega/\Omegaref) L_m \psi -  m \zeta \psi 
= - \ii E\ L_m^2\psi , \label{eq:momentum_adim}
\end{equation}
where we defined  the relative differential rotation 
\begin{equation}
\delta(\theta) =\Omega(\theta)/\Omegaref-1
\end{equation}
 and the operator $L_m$ such that
\begin{equation}
    L_m \psi  =  \frac{1}{\sin\theta} \frac{\id}{\id \theta} \left( \sin\theta \ \frac{\id \psi}{\id \theta}  \right) - \frac{m^2}{\sin^2\theta} \psi,
\end{equation}
i.e. the horizontal Laplacian on the unit sphere (also called the associated Legendre operator).
Equation~(\ref{eq:momentum_adim}) with $E=0$  reduces to  \citet{Watson1981}'s equation when written in a rotating frame.
However, when $E\neq0$, Eq.~(\ref{eq:momentum_adim}) is  fourth-order, which has profound implications for the spectrum.
 Four boundary conditions are required. 
 We impose that the flow vanishes at the poles:
\begin{equation}
\psi(0) = \psi(\pi) = 0 \quad \textrm{and} \quad \frac{\id \psi}{\id\theta} (0) = \frac{\id \psi}{\id\theta} (\pi)=0.
\label{eq:Boundary}
\end{equation}

\begin{figure*}[t]
\centering
\includegraphics[width=0.8\linewidth]{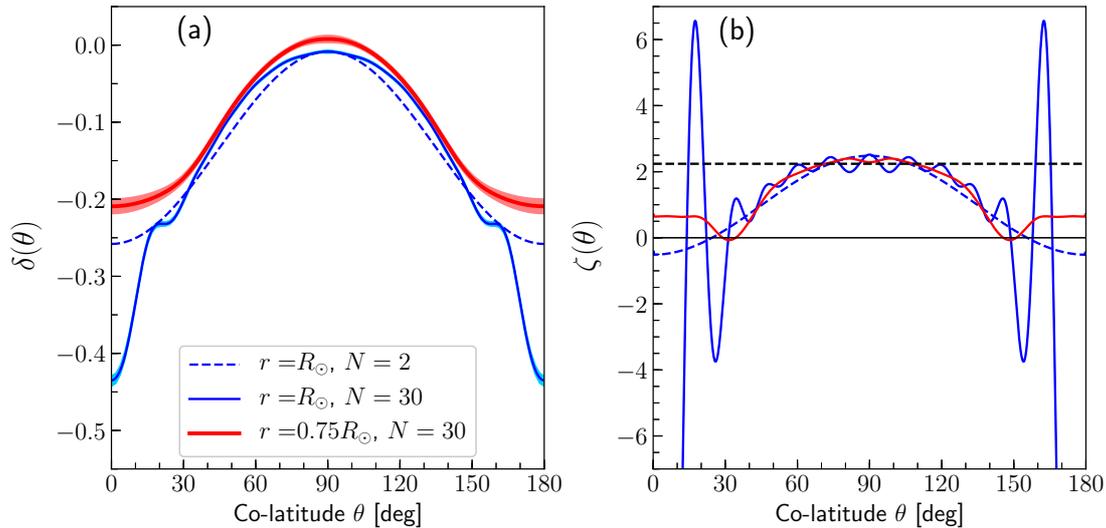}
\caption{(a) Normalized solar differential rotation rate $\delta(\theta)$ at different radii, inferred from helioseismology and  averaged over 1996 to 2018 (the shaded area covers $\pm 3\sigma$), with respect to  the Carrington rotation rate,   $\Omegaref = \Omega_{\rm Carr}$. 
The solid curves show the $N=30$ fits to the solar rotation rates. The dashed curve show the $N=2$ fit to the solar surface rotation rate, i.e. $\Omega_0 + \Omega_2 \cos^2\theta$ with $\Omega_0 / 2\pi = 452$~nHz and $\Omega_2 / 2\pi = -114$~nHz.  
(b) The function $\zeta$ defined by Eq.~\eqref{eq:zeta} is plotted for each of the three cases of panel (a). 
For $r=R_\sun$ and $N=30$, the function reaches a minimum at $\zeta(\theta=0) \approx -22$ (not shown on the plot). The oscillations at the surface are due to the $m=0$ zonal flows (also known as torsional oscillations). 
{The dashed black horizontal line shows $\beta - U_\phi'' \approx 2.24$, where $\beta= 2\Omega_{\rm ref}/r$, which plays the role of $\zeta$ in the equatorial $\beta$ plane, \citep[see equation~(17) in][]{Gizon2020}.}
}
\label{fig:rotation}
\end{figure*} 

\subsection{Numerical solutions}

In the inviscid case ($E =0$ in Eq.~\ref{eq:momentum_adim}), critical latitudes $\theta_c$ appear where $m \delta (\theta_{\rm c}) - \omega/\Omegaref = 0$, i.e. where  \begin{equation}
    \Omega(\theta_c) - \Omega_{\rm ref} = \textrm{Re}[\omega] / m .
\end{equation} The stream eigenfunctions are continuous but not regular at $\theta_c$, thus $\bu'$ is singular there  \citep[see][for the $\beta$-plane problem]{Gizon2020}. 

When including viscosity, the critical latitude is replaced by a viscous layer whose width is proportional to $E^{1/3}$. The eigenfunctions are now regular.
The solutions can be expended onto a series of normalized associated Legendre polynomials up to order $\ell=L$,
\begin{equation}
 \psi(\theta)=\sum_{\ell=m}^{L}  b_{\ell} P_\ell^m(\cos\theta) , \label{eq:exppsi}
\end{equation}
where the $P_\ell^m$ are normalized such that $\int_{-1}^1 [P_\ell^m(\mu)]^2  \mathrm{d}\mu = 1$.
We insert this expansion into Eq.~\eqref{eq:momentum_adim} and project onto a particular polynomial $P_\ell^m$. Using 
\begin{equation}
L_m  P_\ell^m (\cos\theta) = - \ell (\ell +1) P_\ell^m(\cos\theta)    ,
\end{equation}
we obtain the following matrix equation:
\begin{equation}
 \ell (\ell+1) (\omega/ \Omegaref) b  = -  \ii E\ell^2(\ell+1)^2 b + C b \label{eq:Eigenproblem},
\end{equation}
where  $b=[b_m\ b_{m+1} \cdots\  b_L]^T$ is a vector and the matrix $C$ has elements 
\begin{equation}
    C_{\ell \ell'}= m \int_{0}^{\pi} \left[\ell'(\ell'+1) \delta + \zeta \right] P_{\ell}^m(\cos\theta)P_{\ell'}^m(\cos\theta)\sin\theta\ \id\theta. \label{eq:C}
\end{equation}
Equation~(\ref{eq:Eigenproblem}) defines an eigenvalue problem where $\omega$ is the eigenvalue and $b$ the associated eigenvector.
The differential rotation (through $\delta$ and $\zeta$)  couples the different values of $\ell$ and $\ell'$, so that the  matrix $C$ is not diagonal (the eigenfunctions are not $P_\ell^m$). Under the assumption that the rotation profile is symmetric about the equator, the problem decouples into odd and even values of $\ell$ and has to be solved separately for the symmetric and antisymmetric eigenfunctions. {Throughout this paper, we call symmetric the modes that have a north-south symmetric stream function, i.e. that are symmetric in $u'_\theta$ and antisymmetric in $u'_\phi$. The modes with a north-south antisymmetric stream function are called antisymmetric.}
For the sake of simplicity, we do not consider the case of a general rotation profile with North-South asymmetries. This would result in eigenfunctions that are not  symmetric nor antisymmetric.

Thanks to the decomposition given by Eq.~\eqref{eq:exppsi} the boundary conditions, Eqs.~(\ref{eq:Boundary}), are automatically satisfied for $m > 1$. For $m=1$, however, the boundary condition  on the derivative is not satisfied because $\id  P_\ell^1 (\cos\theta) / \id \theta\neq 0$ at the poles. The  modifications for this case are discussed in Appendix~\ref{sect:m1}.

\begin{table*}[h]
    \centering
    \tiny
    \begin{tabular}{llll}
    \hline \hline
    Differential rotation
         &   2D hydrodynamics  & Differential rotation  & Solar observations \\
         (on a sphere, this paper)
         &   (plane Poiseuille flow) &    ($\beta$-plane approximation) & \\
        \hline
        Rossby 
        & --- & \citet[][no viscosity]{Rossby1939}
        & \citet{Loeptien18}
        \\
        High-latitude & A family or  `wall modes' \citep{Mack1976} &  \citet{Gizon2020} & \citet{Gizon2021} \\
        Critical-latitude & P family or `center modes' \citep{Pekeris1948} &  \citet{Gizon2020} & \citet{Gizon2021} \\
        Strongly damped & S family or `damped modes' \citep{Schensted1961} &  \citet{Gizon2020} & ---
    \end{tabular}
    \caption{Correspondence of  terminologies for viscous modes in shear flows. } 
    \label{tab:typeOfModes_hydro_obs}
\end{table*}

\begin{table*}[h]
\tiny
\centering
\begin{tabular}{l l}
\hline \hline 
Modes & Basic properties (all modes are retrograde, $\omega_r<0$)\\
\hline
\multirow{2}{*}{Rossby} & Modes restored by the Coriolis force with frequencies near  $\omega \approx -2m\Omega_0/\ell(\ell+1)$,  where $\ell=m$ for the  \\ & equatorial R mode, $\ell=m+1$ for R$_1$, and $\ell=m+2$ for R$_2$. \\
\hline
\multirow{2}{*}{High-latitude}& Modes whose eigenfunctions have largest amplitudes in the polar regions. Their frequencies  are the most  \\
&  negative in the rotating frame. The least damped mode at fixed $m$ has frequency $\omega \approx m \Omega_2$.  \\
    \hline
\multirow{2}{*}{Critical-latitude}& Modes whose eigenfunctions have the largest amplitudes at mid- or low-latitudes, between their critical \\
& latitudes.  Their frequencies are the smallest in absolute value, with $\textrm{Re}[\omega] \approx \textrm{Im}[\omega]$.  \\
\hline
\multirow{2}{*}{Strongly damped}  & Modes with very large attenuation ($|\textrm{Im}[\omega]| \geq |\textrm{Re}[\omega]|$),  whose eigenfunctions are highly oscillatory  around their \\
& critical latitudes and  frequencies satisfy $\textrm{Re}[\omega] \approx  m \Omega_2/2$  \citep[for the $\beta$-plane analogy, see][]{GRO68}.
\end{tabular}
\caption{Description of the viscous modes discussed in this paper, for the case $\Omega=\Omega_0 + \Omega_2 \cos^2\theta$. 
The high-latitude, critical-latitude, and strongly damped modes owe their existence to the presence of viscous critical layers. Rossby modes would exist even in the case of vanishing differential rotation (i.e. without critical layers).}
\label{tab:typeOfModes}
\end{table*}

\subsection{Input parameters: viscosity and rotation profile} \label{sect:solarProfiles}

The only input quantities required to solve the eigenvalue problem are the viscosity and the rotation profile. 
In this paper, we choose an Ekman number at the surface  $E = 4 \times 10^{-4}$, which means  $\nu = 125$~km$^2$/s  and an associated Reynolds number of 300. This choice leads to  a good match with the observed eigenfunctions of the equatorial Rossby modes \citep{Gizon2020}. The influence of the viscosity on the spectrum and the eigenfunctions is studied in Sect.~\ref{sect:viscosity}.

Regarding rotation, we use the profile inferred by helioseismology from  \citet{Larson2018}. Since the first and second derivatives of $\Omega(r,\theta)$ with respect to $\theta$  are required to compute $\zeta$, we first write the profile as a truncated series of harmonic functions to smooth it:
\begin{equation}
    \Omega(\theta; r,N) =   \sum_{\ell=0}^N \Omega_\ell(r) \cos^\ell\theta . \label{eq:rotationLawCos}
\end{equation}
For each  $r$, the coefficients $\Omega_\ell$, $0\leq \ell \leq N$ are found by fitting the observations 
taking the random errors into account. These coefficients  change with the solar cycle (torsional oscillations). For north-south symmetric rotation, the odd coefficients are zero. Rotation from global-mode helioseismology is north-south symmetric by construction, however general rotation profiles could be used using the methods of the present paper.
Note that the coefficients in this decomposition depend on $N$ because the powers of $\cos\theta$ are not orthogonal. 

The (symmetric) rotation profile from \citet{Larson2018} averaged over 1996--2018  is shown on the left panel of Fig.~\ref{fig:rotation}. Close to the surface, an expansion with $N$  up to $30$ is required to capture the  sharp change of slope of the profile around $60^\circ$ latitude, as well as the solar-cycle related zonal flows. 
Adding more terms would  not change the surface profile significantly.
Below the near-surface shear layer, an expansion with $N=4$ is generally sufficient to capture the rotation profile. The quantity $\zeta$, which involves the first and second derivatives of $\Omega(\theta;r,N)$ with respect to $\theta$ shows strong variations near the surface (right panel of Fig.~\ref{fig:rotation}). These will have an important effect on the spectrum as shown in the next section.

\begin{figure*}[t]
    \centering
    \includegraphics[width=\linewidth]{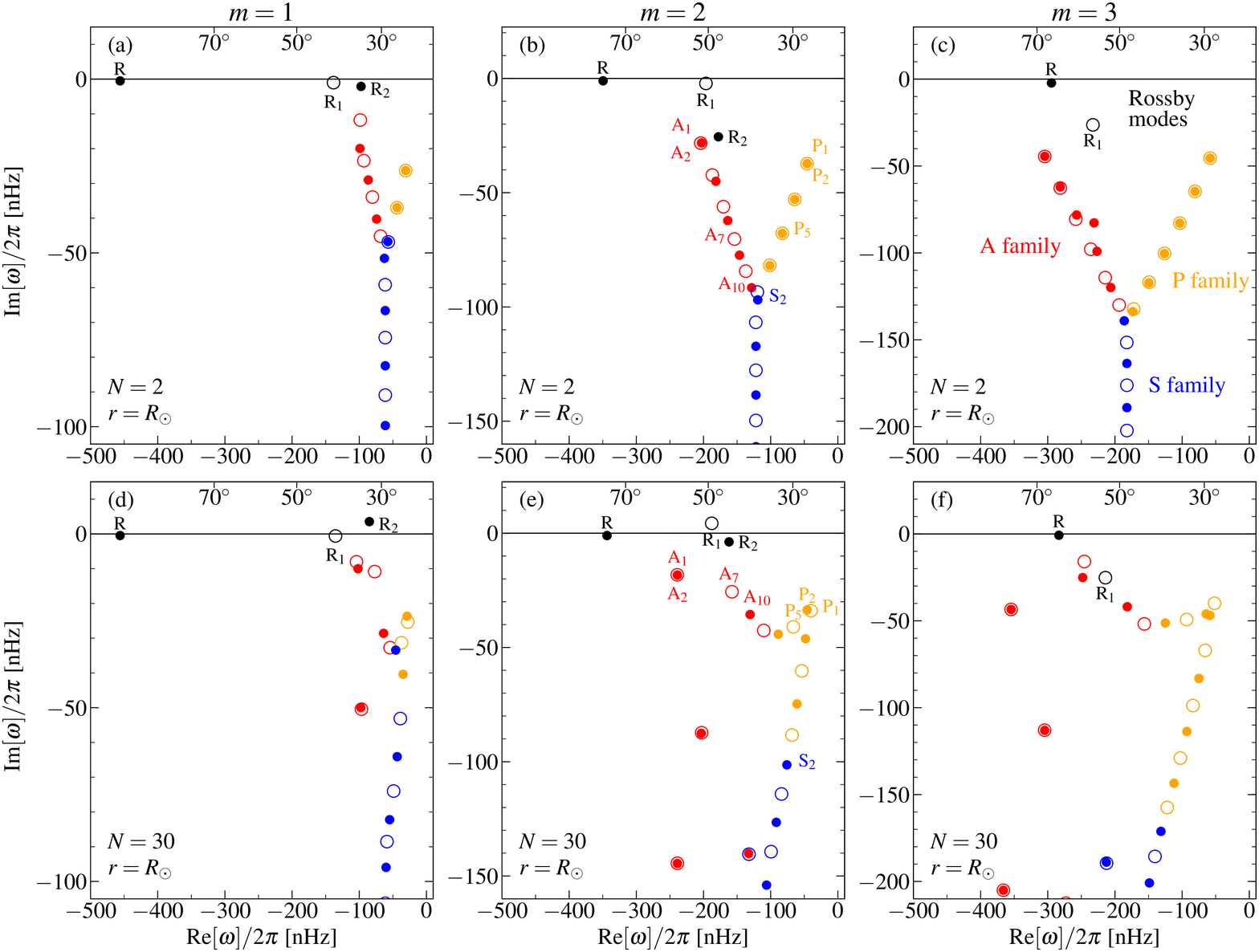}
    \caption{Eigenfrequencies $\omega$  for $m=1$ to 3 (from left to right). The Ekman number is $E=4\times10^{-4}$. 
    The top panels are for  $\Omega(N = 2)=\Omega_0 + \Omega_2 \cos^2\theta$, with $\Omega_0 / 2\pi = 452$~nHz and $\Omega_2 / 2\pi = -114$~nHz and the bottom panels for the solar rotation rate  $\Omega(N = 30)$. Modes with north-south symmetric eigenfunctions are shown with filled symbols, those with antisymmetric eigenfunctions are shown with open symbols. The frequencies are expressed in the Carrington frame of reference.  We note that the imaginary part of the frequency is connected to the mode linewidth  $\Gamma$ through $\Gamma = - 2 \omega_i$.
    The top axis of each plot marks the frequencies corresponding to the critical latitudes  $30^\circ$, $50^\circ$, and $70^\circ$ (this is not a statement about the latitude of maximum power). 
    The three branches for the A, P, and S families  are highlighted with different colors \citep[see][for a comparison with the spectrum of the Orr-Sommerfeld equation]{Mack1976}. The modes denoted R, R$_1$, and R$_2$ are closest to the traditional r-modes with $\ell=m$, $\ell=m+1$, and  $\ell=m+2$ respectively \citep[see, e.g.,][]{Saio1982}. The denomination of the modes in the solar case (bottom panels) is obtained by tracking the modes from the $N=2$ case (top panels). See the evolution between the two profiles on the \href{ https://doi.org/10.17617/3.OM51HE}{online movies} for $m=1$ and $m=2$.
    }
    \label{fig:spectrum_m1_4}
\end{figure*}

\section{Spectrum for simplified rotation laws}

\subsection{Uniform rotation}

In the case of uniform rotation, $\Omega =\Omega_0$, the complex eigenvalues can be obtained directly from Eq.~\eqref{eq:Eigenproblem} (the matrix $C$ is diagonal because  $\delta$ and $\zeta$ are constant):
\begin{equation}
    \omega_{\ell m}^0 =  m(\Omega_0-\Omega_{\rm ref}) -   \frac{2m \Omega_0}{\ell(\ell+1)} - \ii E \Omegaref \ell(\ell+1)
     . \label{eq:eigUniform}
\end{equation}
The modes are classical Rossby modes with eigenfunctions  $P_\ell^m(\cos\theta)$. The modes are stable as their  eigenvalues all have negative imaginary parts. For a given $m$, the sectoral Rossby mode ($\ell = m$) is the least damped mode.

\subsection{Two-term  differential rotation and $E \gtrsim  10^{-4}$} \label{sect:spectrum_cos2}

Here we solve the eigenvalue problem given by Eq.~\eqref{eq:Eigenproblem}  using  $\Omega_0 + \Omega_2 \cos^2\theta$ as an approximation to the solar differential rotation at the surface. Fitting the observed surface rotation profile, we use $\Omega_0 / 2\pi = 452$~nHz and $\Omega_2/2\pi = -114$~nHz. This simple profile allows to make the connection with the study of \citet{Gizon2020}, where a parabolic flow was used in the $\beta-$plane,  and serves as a reference to track and identify the modes when the rotation profile is steeper and more solar-like. We use an Ekman number $E = 4 \times 10^{-4}$ which is close to the value of the eddy viscosity at the solar surface.
As shown in Fig.~\ref{fig:spectrum_m1_4} (top panels), the spectrum takes a Y-shape in the complex plane.  This spectrum is characteristic of the spectrum of the Orr-Sommerfeld equation for a  parabolic (Poiseuille) flow.  The three branches of the spectrum have been called by \citet[][his figure 5]{Mack1976} the P family \citep[after][]{Pekeris1948}, the S family  \citep[after][]{Schensted1961}, and the A family; they correspond respectively to the `center modes', the `damped modes', and the `wall modes'. We refer to the monographs by  \citet{DrazinReidBook} and \citet{SCH02} for more details about this problem in the context of hydrodynamics, as well as to the review paper by \citet{Maslowe2003} for discussions about critical layers in shear flows. The correspondence between the terminologies used in hydrodynamics and in solar physics  is given in Table~\ref{tab:typeOfModes_hydro_obs}. We find that the three branches are still present  for low values of $m$, even though the $\beta$-plane approximation is no longer justified. 
We label the symmetric modes with an even index and  the antisymmetric modes with an odd index. They are ordered on the branches  such that the  index increases with the attenuation (see Fig.~\ref{fig:spectrum_m1_4}, top panel for $m=2$).

In addition to the above modes, up to three other modes are present in the spectrum at low $m$ values. These modes are outside the three branches of the spectrum and are strongly affected by the Coriolis force. These are Rossby modes.  For $m$ large enough, only the equatorial Rossby mode (denoted by the letter R) is easily identifiable outside the branches \citep[as in the $\beta-$plane approximation, see][their figure~4]{Gizon2020}. 
We identify up to two additional Rossby modes  in the frequency spectra (denoted R$_1$ and R$_2$). These can be traced back to the traditional Rossby modes in the case of uniform rotation. By progressively turning on the  differential rotation, i.e. $\Omega(\theta) = \Omega_0 + \epsilon \Omega_2 \cos^2\theta$ with $\epsilon$ increasing from 0 to 1, we find that these correspond to the modes with $\ell=m+1$ and $\ell=m+2$ in the case of uniform rotation  ($\epsilon=0$), see the \href{ https://doi.org/10.17617/3.OM51HE}{online movies} for $m=1$ and $m=2$. The study of Rossby waves in stars and more generally in astrophysics is a broad topic. We refer the reader to the recent review by \citet{ZAQ21}.
Basic properties of the modes studied in this paper are summarized in  Table~\ref{tab:typeOfModes}.

\begin{figure*}[t]
\begin{tabular}{cc}
\includegraphics[height=6.5cm]{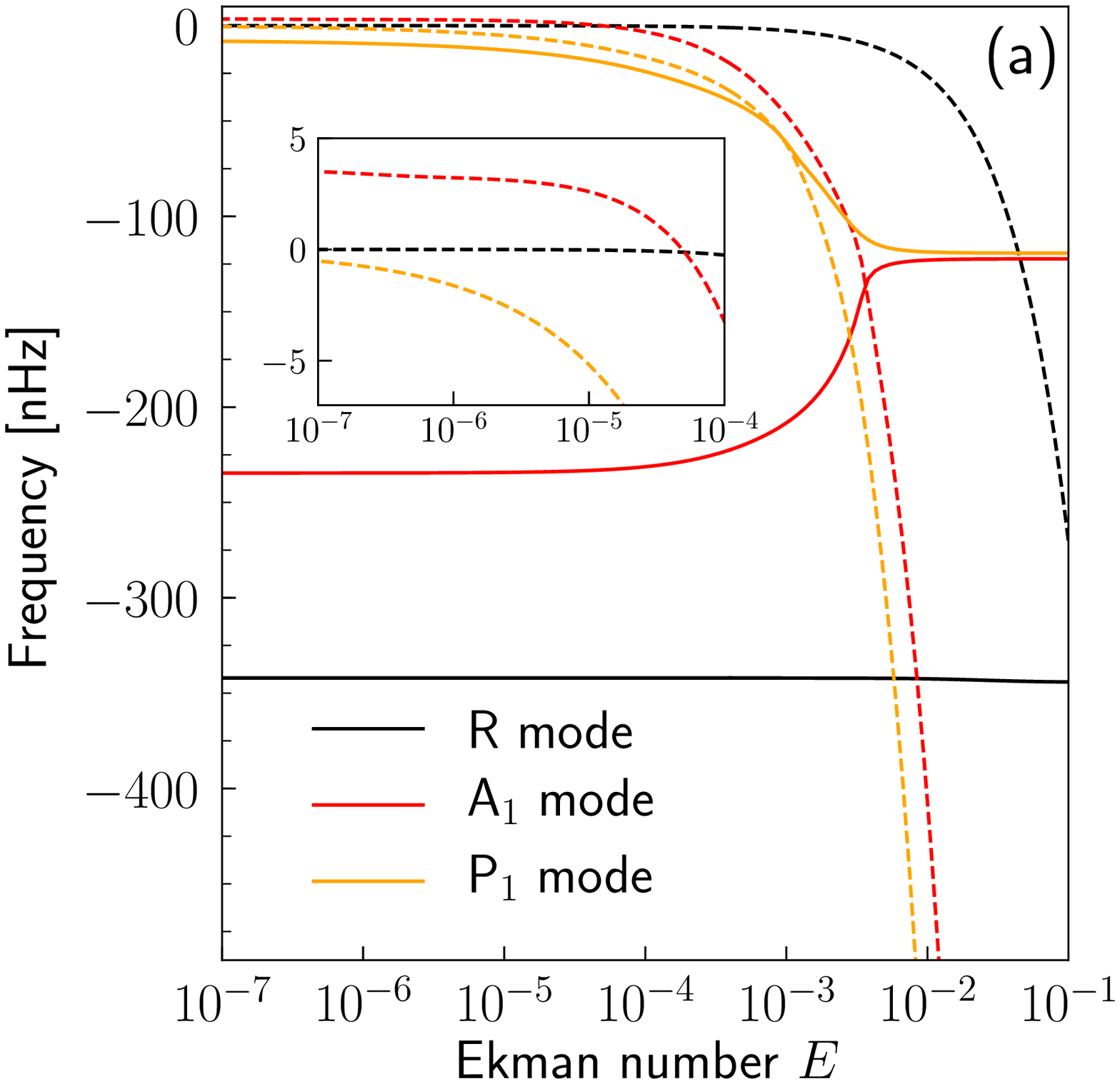}     
&  \includegraphics[height=6.5cm]{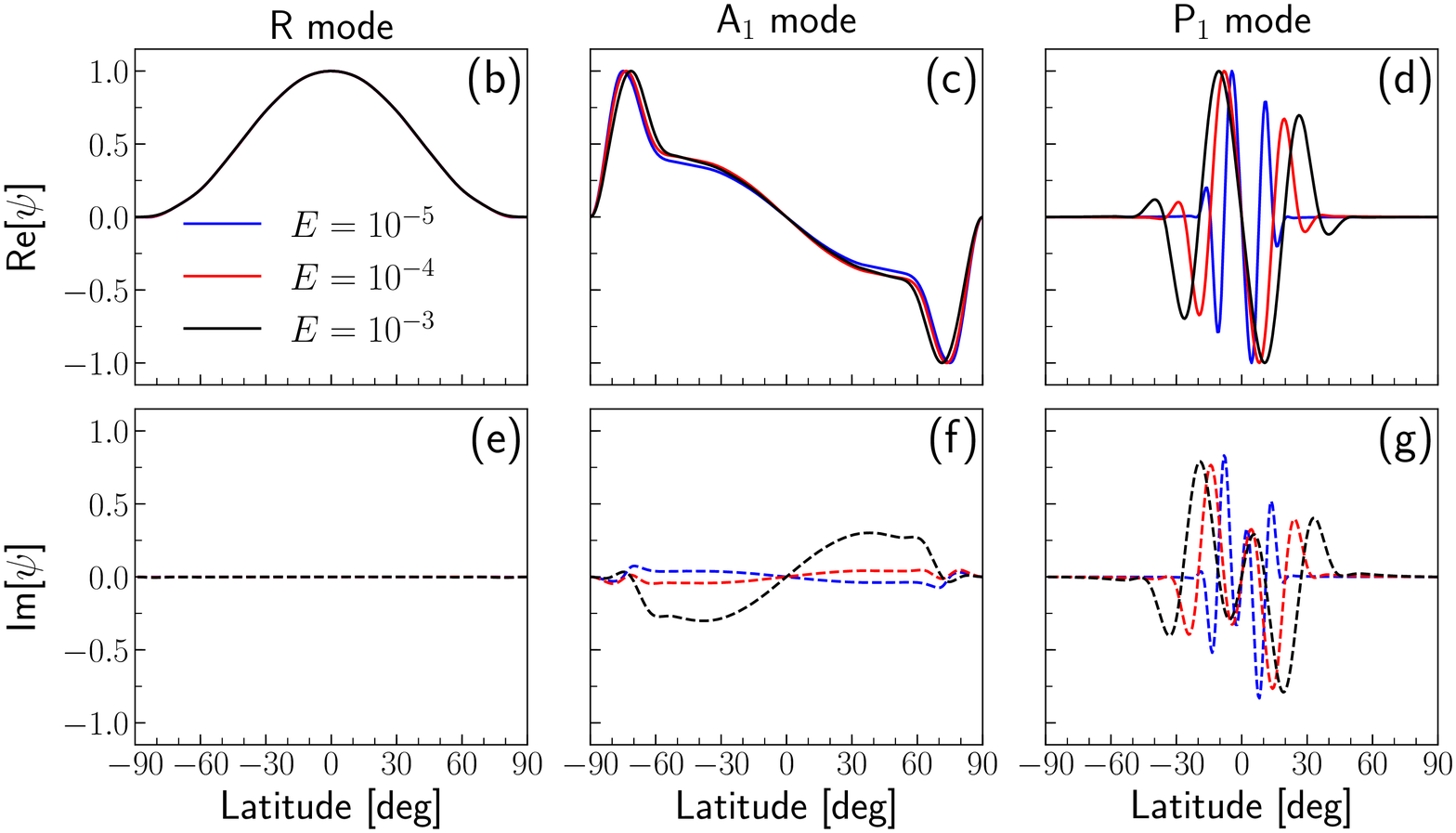}
\end{tabular}
\caption{(a): Real (solid curves) and imaginary (dashed curves) parts of the frequencies of the modes R, A$_1$ and P$_1$ for $m=2$ as a function of  Ekman number, in the case of observed solar surface rotation profile.
Panels (b)-(g) show the real (top panels) and imaginary (bottom panels) parts of the stream functions for the same modes at three selected values of the Ekman number. See also the top panels of Fig.~\ref{fig:eigenfunctions2D_m2} for a 2D representation of the stream functions for these modes.
}
\label{fig:influenceViscosity}
\end{figure*}

\subsection{Two-term  differential rotation and $E \lesssim 10^{-4}$}

While the value of the viscosity is well constrained at the solar surface, its value inside the Sun remains largely unknown. Several models presume that the viscosity should decrease with depth  \citep[see e.g. figure 1 of][]{Munoz2011}. The top panels in Fig.~\ref{fig:spectrum_m1_3_0_75} show the spectrum of the modes $m=1$, 2, and 3 when the Ekman number is set to $E = 2 \times 10^{-5}$ (instead of $4 \times 10^{-4}$ as in Fig.~\ref{fig:spectrum_m1_4}). For $m=2$ and $m=3$, the spectrum is not Y-shaped but more complex as many S modes are now situated on a plateau with nearly constant imaginary part. The two branches corresponding to  the A and P families are still clearly visible.

\section{Spectrum for solar rotation at the surface} \label{sect:surfaceDiffRot}

When we use a very good approximation to the solar differential rotation at the surface ($N=30$ terms in the expansion to describe  $\Omega$), we find 
that the Y-shape of the spectrum is much harder to identify and that many modes  move away from  the branches, see bottom panels in Fig.~\ref{fig:spectrum_m1_4} for $1\le m \le 3$.
To label the modes, we track their frequencies in the complex plane by slowly transitioning from the case $\Omega(N=2)$ to the solar case $\Omega(N=30)$, see, e.g., the \href{ https://doi.org/10.17617/3.OM51HE}{online movie} for $m=2$.

The real parts of the eigenfrequencies of the modes R, R$_1$ and R$_2$ do not change very much compared to the case of the  two-term rotation profile. However the imaginary parts of the eigenfrequencies do change. In particular, those of the (symmetric)   R$_2$ mode with  $m=1$ and the (antisymmetric)   R$_1$ mode with $m=2$ are now positive, i.e. these modes are unstable (self-excited). 
The second derivative of the rotation profile has a strong influence on the damping of the modes, especially the high-latitude and critical-latitude modes (see, e.g,  \href{ https://doi.org/10.17617/3.OM51HE}{online movies}). While the damping rates of the A and P modes are large for $\Omega(N=2)$, some of these modes become less damped for the solar-like $\Omega(N=30)$ and are thus more likely to be observable in the solar data.

Figure~\ref{fig:eigenfunctions} shows the eigenfunctions for selected modes with $m=2$ shown in Fig.~\ref{fig:spectrum_m1_4}.
The three Rossby modes have eigenfunctions that are close to $P_\ell^m$, i.e. to the case of uniform rotation. For $m=2$, the critical latitudes are very close to the poles and have little effect on  the eigenfunctions. For larger values of $m$, the eigenfunctions are confined between the critical layers and have a significant imaginary part \citet[see][for $m=10$]{Gizon2020}.

The eigenfunctions of the high-latitude modes A$_1$ and A$_2$ have a modulus that peaks near the critical layers at $\approx \pm 70^\circ$ and the argument between the real and imaginary parts varies below this latitudes (which implies a spiral pattern there). The eigenfunctions  depend sensitively on the rotation profile. The computed frequencies of these modes ($-224$~nHz) are not very far from the observed frequencies at $-171$ nHz and $-151$~nHz \citep[][their table~1]{Gizon2021}. We will show in Sect.~\ref{sect:modes}, that an improved match can be obtained when the solar rotation profile is taken deep in the convection zone.

The critical-latitude modes P$_1$ and P$_2$ have  eigenfunctions that oscillate around and below the critical latitudes. Their real and imaginary parts are not in phase as functions of latitude  and have roughly the same amplitude.
These modes are not very different from the case of the two-term rotation profile. They resemble the observed $m=2$ critical-latitude modes at frequencies $-12$ nHz and $-24$~nHz reported by \citet{Gizon2021}.

\subsection{Effect of turbulent viscosity} \label{sect:viscosity}

In the uniform rotation case, the viscosity only influences the imaginary part of the eigenfrequencies as shown by Eq.~(\ref{eq:eigUniform}). However, the picture changes when differential rotation is included. The left panel of Fig.~\ref{fig:influenceViscosity} shows the eigenfrequencies of several modes with $m=2$ as a function of the Ekman number when the surface solar differential rotation profile is considered. The imaginary part goes to zero as the viscosity tends to 0 and increases drastically when the Ekman number becomes larger than $10^{-3}$. More surprisingly, the real part is also significantly affected for modes A$_1$ and P$_1$, by an amount  of more than 100~nHz depending on the value of the viscosity. The R mode is little affected for this small value of $m$. For larger values of $m$, we observe a change in the R-mode frequency of, e.g.,  15~nHz for $m=6$ and 30~nHz for $m=10$ over the range of  Ekman numbers covered by the plot. All these variations would be measurable in the solar observations and thus the modes may be used as probes of the viscosity in the solar interior. The eigenfunctions are also affected (see Fig.~\ref{fig:influenceViscosity}) in particular the imaginary part of the mode A$_1$  changes sign with viscosity and the mode P$_1$ becomes more and more confined between the viscous layers as viscosity decreases. The shape of the R-mode  eigenfunction does not depend significantly on the viscosity for this small value of $m$, unlike for  larger values of $m$ as already discussed by \citet{Gizon2020}.

\begin{figure*}[t]
\includegraphics[width=\linewidth]{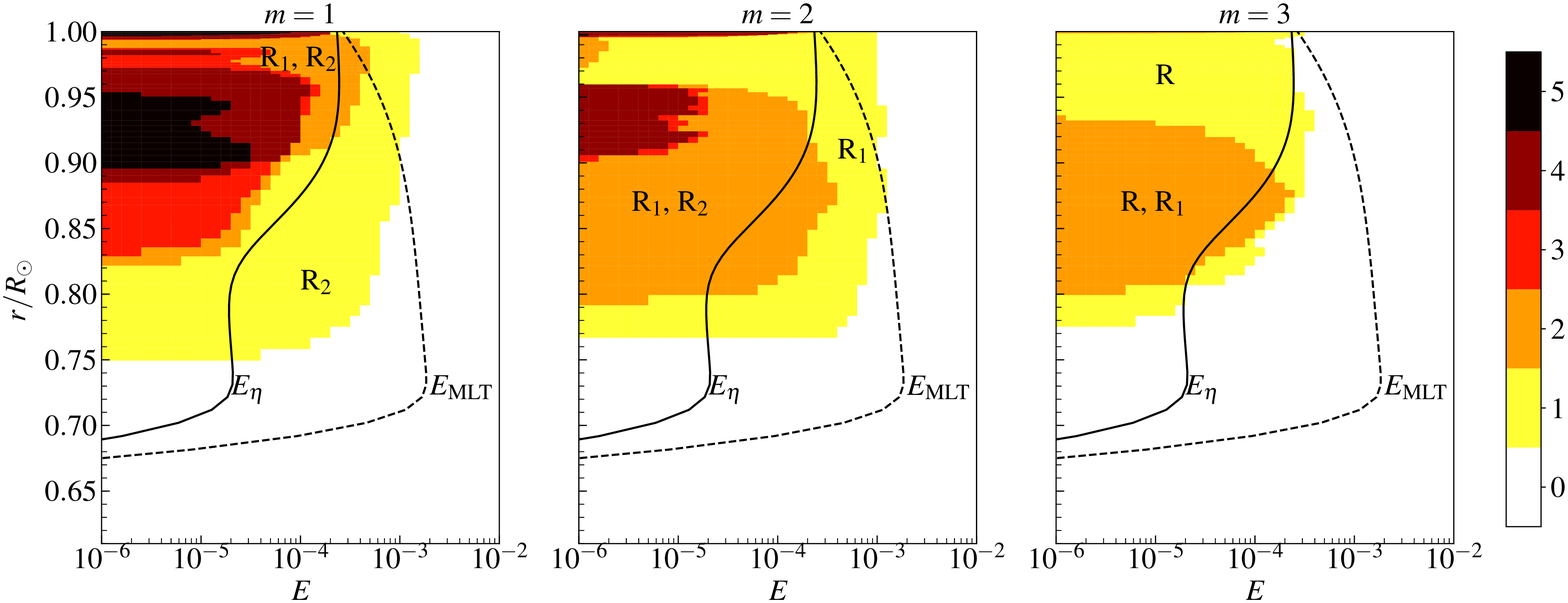}
\caption{Number of unstable modes as a function of the Ekman number at different radii for $m=1,2,3$. For $m \ge 4$ all modes are stable. The solar differential rotation is from \citet{Larson2018}. The dashed black line gives the Ekman number $E_{\rm MLT}$ as a function of radius using  the turbulent viscosity from mixing-length theory  \citep{Munoz2011}.
The solid black line shows the Ekman number $E_\eta$ using the  quenched  diffusivity model  proposed by \citet{Munoz2011}, see their figure~4b. The regions corresponding to the most unstable modes (R$_2$, R$_1$, R) and to the two most unstable modes (R$_2$ and R$_1$, R$_1$ and R$_2$, R and R$_1$) are colored and labelled with these modes. The color bar indicates the total number of unstable modes at any point in the diagram.
}
\label{fig:stability_depth_Ekman}
\end{figure*}

\subsection{Effect of meridional flow}

In order to study the importance of the meridional flow $V(\theta)$, we consider the background flow
\begin{equation}
    \bU=(\Omega - \Omegaref) \ez \times \br + V \etheta.
\end{equation}
At the surface, we take a poleward flow with maximum amplitude $V_{\rm max}=15$~m/s,
\begin{equation}
    V(\theta) = - 1.54 \times V_{\rm max}  \sin^2\theta \sin 2\theta .
    \label{Vmer}
\end{equation}
The derivation and the discretization of this problem are presented in Appendix~\ref{sect:meridionalFlow_app}. The stream function $\psi$ satisfies Eq.~\eqref{eq:momentum_adimMeridional}.

For $m\gtrsim 5$, the effect of the meridional flow on the R mode was already discussed by \citet{Gizon2020} in the $\beta$ plane.
Figure~\ref{fig:eigenfrequencies_mer} shows the entire frequency spectrum with and without the meridional flow in spherical geometry. For $m=2$, the eigenfrequencies  of the  modes R and A$_1$ are shifted by only  a few nanohertz and their eigenfunctions are not significantly affected  (see Fig.~\ref{fig:eigenfunctions_mer}, left and middle panels). An interesting effect caused by the meridional flow is the change of the imaginary part of the frequency of the R$_1$ mode. This mode becomes even more unstable as a consequence of the meridional flow. 
 
We find that the critical-latitude modes are the most affected by the meridional flow. This is not surprising as their eigenfunctions vary fast at mid-latitudes where the meridional flow is  largest. The real parts of the eigenfrequencies of the $m=2$ modes located on the P branch are getting closer to zero when the meridional flow is included,  with a shift of $\sim 10$~nHz. The meridional flow also stretches the eigenfunctions in latitude, see, e.g., the right panel of Fig.~\ref{fig:eigenfunctions_mer} for the P$_1$ mode for $m=2$.

\subsection{Effect of time-varying zonal flows}

To assess the sensitivity of the modes to the details of the rotation profile, we compute the mode frequencies as a function of time over the last two solar cycles.
We use the inferred rotation profile from helioseseismology  \citep{Larson2018}, obtained at a cadence of  72~days. 
The rotation profiles are averaged over five bins, to obtain yearly averages  during  1996--2018.  The  time variations of the frequencies of the least-damped modes with $m=2$ are plotted in Fig.~\ref{fig:eigenfrequencyWithTime}. 
We find that the frequency of the R mode varies by  less than 1~nHz, in agreement with the calculation of \cite{Goddard2020} using first-order perturbation theory. The critial-latitude mode P$_1$ also changes by a very small amount of  $\pm 2$ nHz.
The frequency of the high-latitude mode A$_1$ varies by up to $\pm 7$~nHz and shows  a 22-year periodicity.  Other modes have frequencies that  strongly vary (such as P$_5$, not shown on the plot, which varies by up to $\pm 10$~nHz).
Since the frequency resolution corresponding to a 3 year time series is $\approx 10$~nHz and the typical linewidth of a mode is also $\approx 10$ nHz,  the frequency shifts of some of the modes may be detectable in observed  time series.

\section{Spectrum for solar rotation at  different radii} \label{sect:depth}

In the previous section we only considered the rotation profile on the solar surface. Here, we consider latitudinal differential rotation at different depths in the solar interior. We study mode stability as a function of depth, and how the dispersion relations of the different modes are affected.

\subsection{Stability analysis} \label{sect:stability}

The  hydrodynamical stability of solar latitudinal differential rotation was  studied in two dimensions for purely toroidal disturbances in the inviscid case. A necessary condition for instability is that the latitudinal gradient of vorticity ($\zeta$, see Eq.~(\ref{eq:zeta})) must change sign at least once with latitude \citep{Rayleigh1879}. \citet{FJO50} proved a more restrictive condition for instability: there exist a $\theta$ and a $\theta_0$ (where $\theta_0$ is a  zero of $\vort$) such that $[\Omega(\cos\theta)-\Omega(\cos\theta_0)] \vort(\theta) < 0$.
As seen in Fig.~\ref{fig:rotation}b, the function $\vort(\theta)$ switches sign multiple times at the surface,  and thus unstable modes could exist. 
In the special case of a rotation law $\Omega = \Omega_0 + \Omega_2\cos^2\theta$, \citet{Watson1981} showed that a necessary condition for  stability is  $-(2/7) \Omega_0 <\Omega_2 <  1.14 \Omega_0$. This condition is met for two-term fits to the solar rotation profile.
\citet{Charbonneau1999} added a third term in $\Omega_4\cos^4\theta $  and found numerically that when $\Omega_4\approx \Omega_2 \approx -0.1 \Omega_0$ (as is the case for the Sun) then two modes (one symmetric and one antisymmetric) are unstable for each of the cases  $m=1$ and $m=2$. They considered different solar rotation profiles inferred from the LOWL, GONG, and MDI p-mode splittings and performed a stability  analysis on spheres at different depths. They found  that all modes are stable below $\approx 0.74 R_\odot$ while some modes with $m=1$ and $m=2$ become unstable in the upper convection zone.

\begin{figure*}[t]
\centering
\includegraphics[width=\linewidth]{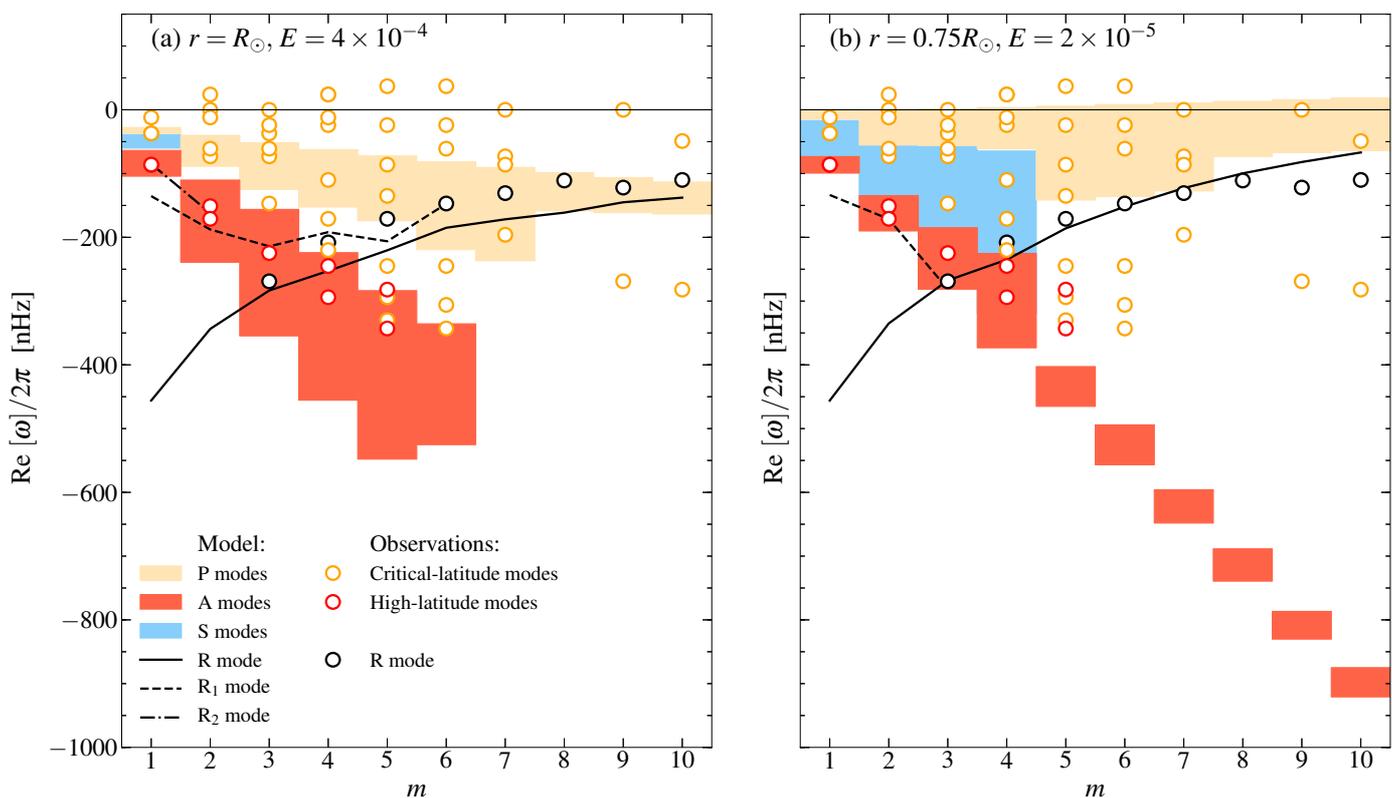} 
\caption{Model dispersion relations for all the modes with  $-\textrm{Im}[\omega]/2\pi<100$~nHz. Shown are the equatorial Rossby modes (black curves for R, R$_1$ and $R_2$), the high-latitude modes (red area), the strongly damped modes (blue area), and the critical-latitude modes (orange area). The differential rotation is solar ($N=30$), both at the surface  ($r=R_\odot$ and $E=4 \times 10^{-4}$, left panel) and at the bottom of the convection zone ($r=0.75R_\odot$ and $E=2 \times 10^{-5}$, right panel). Frequencies are expressed in the Carrington frame of reference. The symbols show the observed modes reported by \citet{Gizon2021}.
}
\label{fig:dispersionRelations}
\end{figure*}

We extended the stability analysis of  \citet{Charbonneau1999} by including viscosity and using  the latest rotation profile from helioseismology \citep{Larson2018}.
Figure~\ref{fig:stability_depth_Ekman} shows the number of unstable modes as a function of the Ekman number and depth  for $m=1$, $2$ and $3$. For $m \ge 4$, all modes are stable. Several modes can be unstable, with up to 12 modes in a small layer below the solar surface for $E < 10^{-4}$ (5 modes with $m=1$, 5 modes with $m=2$, and 2 modes with $m=3$). We find that for $E < 10^{-4}$,  the radius above which some modes become unstable ($\approx 0.75 R_\odot$ for $m=1$, $\approx 0.77 R_\odot$ for $m=2$, and $\approx 0.78 R_\odot$ for $m=3$) does not depend sensitively on viscosity and is almost given by the $E=0$ limit. Only values $E>10^{-4}$ have a stabilising effect. For $E > 5 \times 10^{-3}$ all depths are stable. On the same figure we draw estimates of the Ekman number with depth, either under mixing-length theory or using a quenched diffusivity \citep{Munoz2011}. Using the value $E_\eta$ at each depth, six modes are unstable, two for each value of $m$. These are the modes R$_1$ and R$_2$ for $m=1$ and $m=2$, and R and R$_1$ for $m=3$. Their eigenfunctions at the surface are shown in Fig.~\ref{fig:eigenfunctions2D_unstable}. Since the critical latitude is very close to the pole, it does not affect much the shape of the eigenfunctions which resemble the classical $P_\ell^m$ with $\ell = m, m+1, m+2$ obtained in the case of uniform rotation.
Using the values of $E_{\rm MLT}$, only three modes are unstable (R$_1$ and R$_2$ for $m=1$, and R$_1$ for $m=2$) and no modes are unstable below $0.91 R_\odot$ instead of $0.75 R_\odot$ using $E_\eta$.

The amplitudes of the  inertial modes may thus have diagnostic value to learn about eddy viscosity in the solar interior.

\subsection{Propagation diagram and comparison with observations} \label{sect:modes}

Figure~\ref{fig:dispersionRelations} shows the propagation diagram  for different modes when the radius is either $r =  R_\odot$ or $r =  0.75R_\odot$. We consider  all the modes in the model whose imaginary frequencies are  less  than $100$~nHz (in absolute value) and which would thus be easier to detect on the Sun. Since  several modes of each family are present for each value of $m$, we draw areas in the propagation diagram where these modes are present. 
The observed frequencies  reported by \citet{Gizon2021} are overplotted. We see that observed  high-latitude modes have frequencies that overlap with the region between the slowest and fastest high-latitude modes of the 2D model and are best approximated when the rotation profile is taken at the base of the convection zone. This is not surprising as these modes have most of their kinetic energy there according to 3D modeling \citep[see][]{Gizon2021, Bekki2022}. The observed high-latitude modes for $m=1$, 2, and 3 have frequencies close to the modes R$_1$ and R$_2$ that are self-excited. This may explain why the observed high-latitude modes have the largest amplitudes. However, the eigenfunctions from the model differ from the observations. It is thus difficult to clearly identify these modes among the modes A$_1$, A$_2$, R$_1$, and R$_2$. 
The observed critical-latitude modes with frequencies between $-150$~nHz and 0~nHz may be associated with the dense spectrum of critical-latitude modes. However, the model does not have any mode with positive frequency (in the Carrington frame). The model is also unable to explain some of the observed critical-latitude modes with $\textrm{Re}[\omega]/2\pi< -200$ nHz (for example the modes with $m=9$ and 10 and frequencies around $- 280$~nHz). Figure~\ref{fig:dispersionRelations_critical} shows a different representation of the propagation diagram where the frequency is divided by the wavenumber $m$. 
Modes with similar values of ${\rm Re} [\omega]/m$ have similar critical latitudes for all values of $m$. This diagram highlights the separation in latitude between the different families of modes.

Figure~\ref{fig:eigenfunctions2D_m2} is a comparison of the eigenfunctions of the modes R, P$_1$ and A$_1$ calculated at the surface and at the bottom of the convection zone for $m=2$. For this low $m$ value, the R-mode eigenfunction changes little with depth. This suggests that the 2D problem captures the essential physics of R modes and that the assumption that these modes are quasi-toroidal is well justified. This conclusion is confirmed by solving the 3D eigenvalue problem in a spherical shell: \citet{Gizon2021} and \citet{Bekki2022} find that the radial velocity of R modes is about two orders of magnitude smaller than their horizontal velocity. On the other hand, the high-latitude mode A$_1$ and (to a lesser extent) the critical-latitude mode P$_1$  have eigenfunctions that change significantly between the surface and the base of the convection zone (middle panels in Fig.~\ref{fig:eigenfunctions2D_m2}).
In the case of the high-latitude modes, this is not too surprising 
\citep[the 2D approximation is poor close to the axis of symmetry, see, e.g.,][]{Rieutord02}. A better understanding of the high-latitude modes requires not only a 3D  geometrical setup, but also the inclusion of a   latitudinal entropy gradient \citep{Gizon2021, Bekki2022}.

\section{Discussion} \label{sect:conclusion}

We extended the work of \citet{Gizon2020} in the equatorial $\beta$ plane to a spherical geometry, in order to study the effects of differential rotation and viscosity on the modes with the lowest $m$ values.
Like in the equatorial $\beta$ plane, viscous critical layers appear and the spectrum contains different families of modes due to the latitudinal differential rotation (Fig.~\ref{fig:spectrum_m1_4}). The high-latitude, critical-latitude, and strongly damped modes \citep{Mack1976} are still present when a realistic solar  rotation profile is used. 
Due to the Coriolis force, the Rossby equatorial modes (R modes) are also present in this problem, like in the $\beta$ plane problem. Using the surface solar rotation profile, two additional Rossby modes, R$_1$ (for $m \leq 6$) and R$_2$ (for $m \leq 2$), are also present. 
Remarkably, the calculated spectrum  resembles closely that of  the inertial modes observed on the Sun by \citet{Gizon2021}, especially when the differential rotation is taken deep in the convection zone, see Fig.~\ref{fig:dispersionRelations}b. The frequencies of some modes in the model are sensitive to the value of the viscosity ($E_{\rm MLT}$ versus $E_\eta$) and could be a good probe of this parameter (Fig.~\ref{fig:influenceViscosity}, left panel). We also find that the zonal flows (Fig.~\ref{fig:eigenfrequencyWithTime}) and the meridional flow (Fig.~\ref{fig:eigenfrequencies_mer}) can affect the modes.
The 2D model presented here is useful to understand the basic physics of toroidal modes on the Sun, however a 2D model  cannot replace a more sophisticated 3D model that includes radial motions, realistic stratification, superadiabaticity (i.e. the strength of the convective driving), and entropy gradients \citep[for a treatment of these effects, see][]{Bekki2022}.

We confirm the result by \citet{Charbonneau1999} that some modes are self-excited, even when viscosity and realistic solar differential rotation are included. The angular velocity  terms in $\Omega_{2p} \cos^{2p} \theta$ with $p\geq2$  are essential to destabilize the system. For the Sun, an oversimplified fit of the form $\Omega_0 + \Omega_2\cos^2\theta$ would lead to the wrong conclusion that the system is  stable.
Using a value of the viscosity corresponding to the quenched diffusivity model from \citet{Munoz2011}, we find that six modes are self-excited. These are the Rossby modes R$_1$ and R$_2$ for $m=1$ and $m=2$, and R and R$_1$ for $m=3$. If the modes R$_1$ and R$_2$ correspond to the high-latitude modes observed by \citet{Gizon2021}, it is understandable why they should have the largest amplitudes. Above $r=0.78 R_\odot$, the $m=3$ equatorial R mode is unstable; this mode is the R mode with the lowest $m$ value that is observed on the Sun. The excitation and damping by turbulent convection of the subcritical modes  will be studied in upcoming papers.

\begin{table}[t]
\caption{
\label{Table:stars}
Rotational stability of selected {\it Kepler} stars for which $\Omega_0$ and $\Omega_2$ have been measured using asteroseismology \citep{Benomar2018}. The $^*$ denotes stars that are unstable by more than one standard deviation ($\sigma$) according to \cite{Watson1981}'s criterion.
} 
\centering
 \begin{tabular}{l  c c c c} 
 \hline \hline
  KIC \#  &  $\Omega_0/2\pi$   & $\Omega_2/\Omega_0 \pm \sigma$ & $\Omega_2/\Omega_0 + 2/7  $ \\ 
  &   (nHz)  & &\\ [0.5ex] 
 \hline
 5184732
 &  $785\pm276$ & $-1.43 \pm 1.87$&
 $-0.6\ \sigma$ \\ 
  6225718
 &  $1725\pm348$ &  $-1.58\pm0.77$ & \textrm{$-2.1\ \sigma$\  (*)}\\
  7510397
 &  $2754\pm469$ &  $-2.11\pm0.92$ & \textrm{$-2.0\ \sigma$\  (*)} \\
  8006161
 &  $722\pm137$ &  $-1.08\pm0.67$ & \textrm{$-1.2\ \sigma$\  (*)} \\
  8379927
 &  $1550\pm230$  & $-1.17\pm0.74$ & \textrm{$-1.2\ \sigma$\  (*)} \\ 
   8694723
 &  $2276\pm324$ &  $-1.23\pm0.61$ & \textrm{$-1.5\ \sigma$\  (*)} \\
   9025370
 & $1015\pm460$ &  $-2.63\pm3.00$ & $-0.8\ \sigma$ \\
  9139151
 &  $1693\pm426$ &  $-2.10\pm1.51$ & \textrm{$-1.2\ \sigma$\  (*)} \\
  9955598
 &  $584\pm243$ &  $-1.72\pm2.11$ & $-0.7 \ \sigma$ \\
  9965715
 & $2321\pm338$ &  $-0.73\pm0.64$ & $-0.7 \ \sigma$ \\
  10068307
 &  $1072\pm313$ &  $-0.92\pm1.09$ & $-0.6 \ \sigma$ \\
  10963065
 &  $1140\pm219$ &  $-0.5\pm0.6$ & $-0.4 \ \sigma$ \\
  12258514
 & $1087\pm432$ &  $-1.3\pm1.4$ & $-0.7 \ \sigma$ \\
 \hline
\end{tabular}
\tablefoot{\citet{Benomar2018} measured the first two coefficients in the expansion $\Omega(\theta) = \alpha_0 + \alpha_2 P_3^1(\cos \theta)/\sin\theta + \cdots$ for a selection of {\it Kepler} stars. Keeping these two terms only,  we have
$\Omega =\Omega_0 + \Omega_2 \cos^2\theta$ with
$\Omega_0 = \alpha_0 - {3\alpha_2}/{2}$
and $\Omega_2 = {15\alpha_2}/{2}$.}
\end{table}

In the case of distant stars, latitudinal differential rotation is detectable with  asteroseismology only for a few \textit{Kepler} Sun-like stars \citep{Benomar2018}. For these stars, the $\Omega_2\cos^2\theta$ term is very large compared to the equatorial value $\Omega_0$, and they are unstable according to Watson's criterion (Table~\ref{Table:stars}). Unfortunately we do not have any information about the high-order terms that fully determine the rotation profile of these stars.

\begin{acknowledgements}
 This work is supported by the ERC Synergy Grant WHOLE~SUN \#810218 and by the DFG Collaborative Research Center SFB 1456 (project C04). L.G. acknowledges NYUAD Institute Grant G1502. LH acknowledges an internship agreement between \mbox{SUPAERO} and the MPS as part of her Bachelor thesis. The source code is available at \href{https://doi.org/10.17617/3.OM51HE}{https://doi.org/10.17617/3.OM51HE}.
\end{acknowledgements}

\bibliography{biblio}{}
\bibliographystyle{aa}

\clearpage
\newpage

\begin{appendix}

\section{Eigenvalue problem for $m=1$} \label{sect:m1}

In order to ensure that $\psi$ satisfies the boundary conditions (Eq.~(\ref{eq:Boundary})) for $m=1$, we expand the solution on a basis of  associated Legendre polynomials $P_\ell^2$ instead of $P_\ell^1$:
\begin{equation}
    \Psi(\theta,\phi,t)=\sum_{\ell=2}^{L}b_\ell P_\ell^2(\cos\theta) \exp(\ii \phi-\ii \omega t) .
\end{equation}
Next, we need to specify how the operators $L_1$ and $(L_1)^2$ act on $P_\ell^2(\cos\theta)$:
\begin{align}
    L_1 P_\ell^2 (\cos\theta)  = & \left( - \ell(\ell+1) + \frac{3}{\sin^2\theta} \right) P_\ell^2(\cos\theta) ,
    \\
    L_1^2 P_\ell^2 (\cos\theta) =&  - \frac{12\cos\theta}{\sin^3\theta} \partial_\theta  P^2_{\ell}(\cos\theta)  
      \nonumber \\
    & \hspace*{-0.5cm} + \left( \ell^2(\ell+1)^2 - \frac{6\ell(\ell+1)}{\sin^2\theta} +\frac{15+6\cos^2\theta}{\sin^4\theta} \right) P^2_{\ell}  (\cos\theta)    .
\end{align}
The eigenvalue problem, Eq.~\eqref{eq:momentum_adim}, becomes
\begin{equation}
\omega/\Omegaref [\ell (\ell+1) I + A] b =  (C+D+G) b -  \ii E\ell^2(\ell+1)^2 b ,
\end{equation}
where  $I$ is the $(L-1) \times (L-1)$ identity matrix, $C$ is defined by Eq.~\eqref{eq:C} with $m=1$, and the elements of the matrices $A$, $D$ and $G$ are given by:
\begin{align}
    &A_{\ell \ell'}= - \int_{0}^{\pi}\frac{3}{\sin^2\theta}P_{\ell'}^2(\cos\theta)P_{\ell}^2(\cos\theta)\sin\theta d\theta, \label{eq:matrixA} \\
    &D_{\ell \ell'}= \int_{0}^{\pi} J_{\ell'}(\theta) P_{\ell'}^2(\cos\theta)P_{\ell}^2(\cos\theta)\sin\theta d\theta, \label{eq:matrixD} \\
    &G_{\ell \ell'}= 12 \ii E \int_{0}^{\pi} \frac{\cos\theta}{\sin^3\theta}\partial_\theta (P^2_{\ell'}(\cos\theta))P_{\ell}^2(\cos\theta)\sin\theta d\theta, \label{eq:matrixG}
\end{align}
with
\begin{equation}
 J_{\ell'}(\theta) =   -\frac{3\delta(\theta)}{\sin^2\theta}
 - \ii E\left(-\frac{6\ell'(\ell'+1)}{\sin^2\theta} +\frac{15+6\cos^2\theta}{\sin^4\theta} \right).
\end{equation}

\section{Inclusion of meridional flow} \label{sect:meridionalFlow_app}

Here we derive the equation for the stream function when the background flow includes both the differential rotation and a meridional flow $V(\theta)$, i.e.
\begin{equation}
    \bU(\theta)=[\Omega(\theta) - \Omegaref] \ez \times \br + V(\theta) \etheta.
\end{equation}
The horizontal components of Eq.~\eqref{eq:Momentum1}  become
\begin{align}
  D_t  u'_\theta - 2 \Omega \cos\theta\ u'_\phi +\frac{V}{r} \partial_\theta u'_\theta + \frac{\partial_\theta V}{r} u'_\theta  &=  -\frac{\partial_\theta \Pi'}{ r} + \nu  \Delta u'_\theta, \label{eq:momentumThMeridional} \\
  D_t  u'_\phi  + \frac{1}{\sin\theta} \frac{\id}{\id\theta} \left( \Omega\sin^2 \theta \right)  u'_\theta
 +  \frac{V}{r\sin\theta} & \partial_\theta (\sin\theta u'_\phi)  \nonumber \\
& =  - \frac{\partial_\phi \Pi'}{r\sin \theta}  + \nu  \Delta u'_\phi. \label{eq:momentumPhiMeridional}
\end{align}
Combining Eq.~\eqref{eq:momentumThMeridional} and Eq.~\eqref{eq:momentumPhiMeridional} and using the definition of the stream function (Eq.~\eqref{eq: u'}), we obtain
\begin{equation}
D_t \Delta \Psi - \zeta \Omegaref  \partial_\phi \Psi /r^2 + \frac{1}{r\sin\theta} \partial_\theta \left( V \sin\theta \Delta\Psi \right)
= \nu \Delta^2\Psi, \label{eq:momentumPsiMeridional}
\end{equation}
where $\zeta(\theta)$ is defined by Eq.~\eqref{eq:zeta}.
For each longitudinal mode, we have  
\begin{equation}
(m \delta  - \omega / \Omegaref) L_m \psi -  m \zeta \psi - \frac{\ii}{\sin\theta} \partial_\theta \left( \tilde{V} \sin\theta\ L_m \psi \right)
= - \ii E L_m^2\psi, \label{eq:momentum_adimMeridional}
\end{equation}
where $\tilde{V} = V / (r \Omegaref)$.

\subsection{Case $m \geq 2$}

In this case, the function $\psi(\theta)$ is expanded according to Eq.~\eqref{eq:exppsi}.
Using   $L_m P_l^m ( \cos\theta)  = - \ell(\ell+1) P_l^m ( \cos\theta)$, we obtain
\begin{align}
  \ell (\ell+1) \omega / \Omegaref \ \psi  =  \bigg( & m \ell(\ell+1) \delta + m \zeta - \ii \frac{\ell(\ell+1)}{\sin\theta} \partial_\theta ( \tilde{V} \sin\theta \, \psi )  \nonumber \\
 & 
- \ii E \ell^2 (\ell+1)^2 \bigg)\
 \psi.
   \label{eq:momentumFinalMeridional}
\end{align}
This leads to the linear system
\begin{equation}
\omega / \Omegaref \ell (\ell+1) b =  (C+C^{\rm mer}) b -  \ii E\ell^2(\ell+1)^2 b
\label{eq:Eigenproblem_Mer},
\end{equation}
where  the matrix $C$ is given by Eq.~\eqref{eq:C} 
and the matrix $C^{\rm mer}$  has elements
\begin{equation}
    C^{\rm{mer}}_{\ell\ell'} = \ii \ell'(\ell'+1) \int_{0}^{\pi}  \tilde{V} \,  \partial_\theta [P_{\ell}^m(\cos\theta)] \,  P_{\ell'}^m(\cos\theta) \sin\theta \mathrm{d}\theta.
\end{equation}
For the meridional flow defined by Eq.~(\ref{Vmer}), the effects of the matrix $C^{\rm{mer}}$ on the spectrum can be seen in Fig.~\ref{fig:eigenfrequencies_mer} for the case $m=2$. Example eigenfunctions are plotted  in Fig.~\ref{fig:eigenfunctions_mer}. 

\subsection{Case $m=1$}

When $m=1$, we use a decomposition of $\psi$ on a basis of $P_\ell^2$ in order to enforce the boundary conditions (see Sect.~\ref{sect:m1}).
The eigenvalue problem becomes
\begin{align}
\omega / \Omegaref \left[ \ell (\ell+1) I + A \right] b =& (C+C^{\rm mer}+D+D^{\rm mer}+G+G^{\rm mer}) b \nonumber \\
& -  \ii E\ell^2(\ell+1)^2 b   \label{eq:EigenproblemM1_mer},
\end{align}
where $D$ and $G$ are defined by Eqs.~\eqref{eq:matrixD}~and~\eqref{eq:matrixG} and
$C^{\rm mer}$ and $D^{\rm mer}$  are given by
\begin{align}
    D^{\rm mer}_{\ell \ell'} &=  3 \ii \int_{0}^{\pi} \frac{ \tilde{V} \cos\theta}{\sin^3\theta}  P_{\ell}^2(\cos\theta)P_{\ell'}^2(\cos\theta) \sin\theta \mathrm{d}\theta, \\
    G^{\rm mer}_{\ell \ell'} &= -3 \ii \int_{0}^{\pi}  \frac{\tilde{V}}{\sin^2\theta}   \partial_\theta [P_{\ell}^2(\cos\theta) ] \  P_{\ell'}^2(\cos\theta)  \sin\theta \mathrm{d}\theta.
\end{align}

\section{Supplementary figures}
\onecolumn

\begin{figure*}[t]
    \centering
    \includegraphics[width=\linewidth]{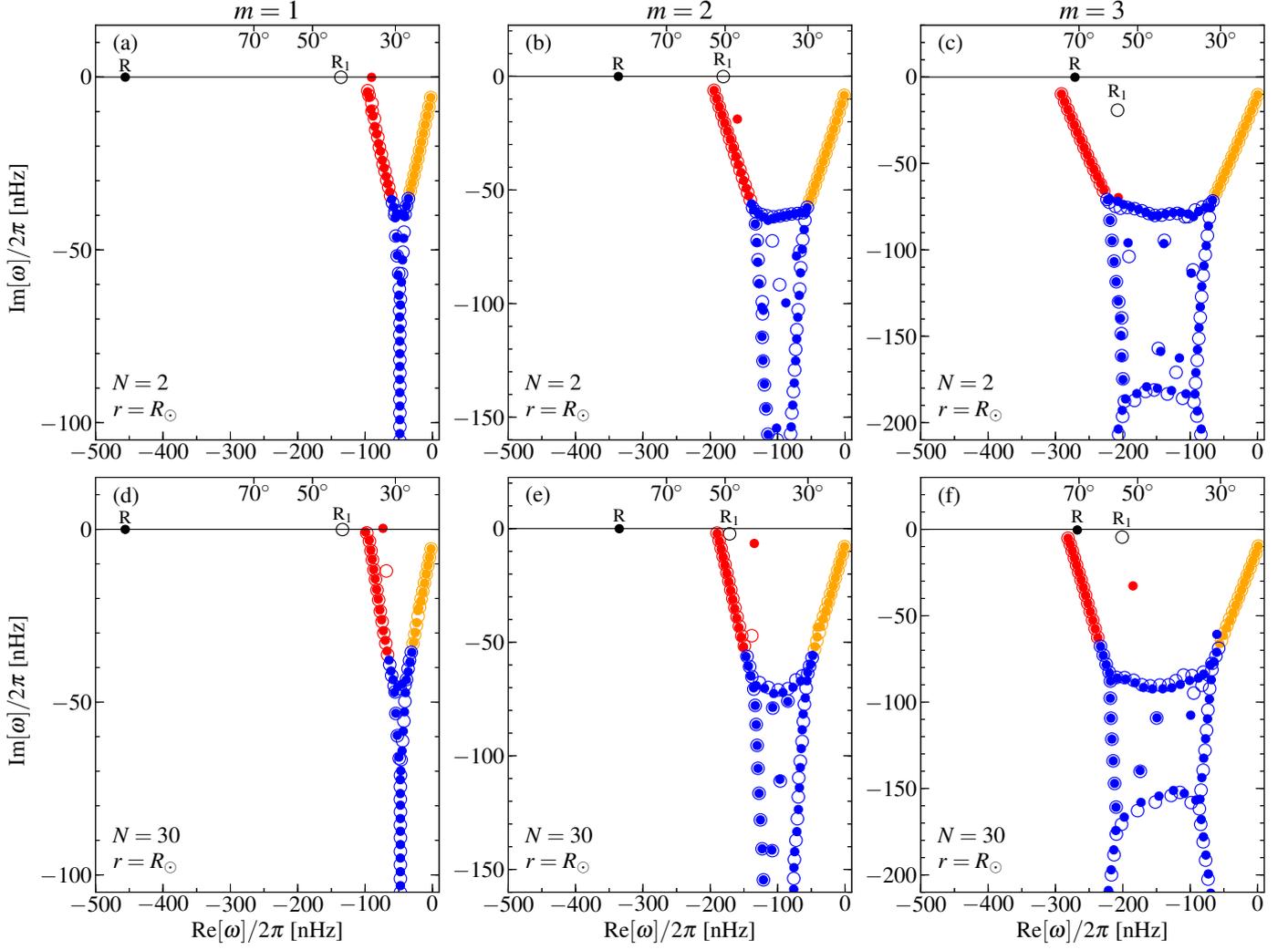}
    \caption{Eigenfrequencies $\omega$  for $m=1$ to 3 (from left to right) when the Ekman number is $E=2\times10^{-5}$. 
    The labels and the colors are the same as in Fig.~\ref{fig:spectrum_m1_4}. The top panels are for  $\Omega(N = 2)=\Omega_0 + \Omega_2 \cos^2\theta$ and the bottom panels for the solar rotation rate  $\Omega(N = 30)$ at a radius $r = 0.75 R_\odot$.
    }
    \label{fig:spectrum_m1_3_0_75}
\end{figure*}

\begin{figure*}
\centering
\includegraphics[width=\linewidth]{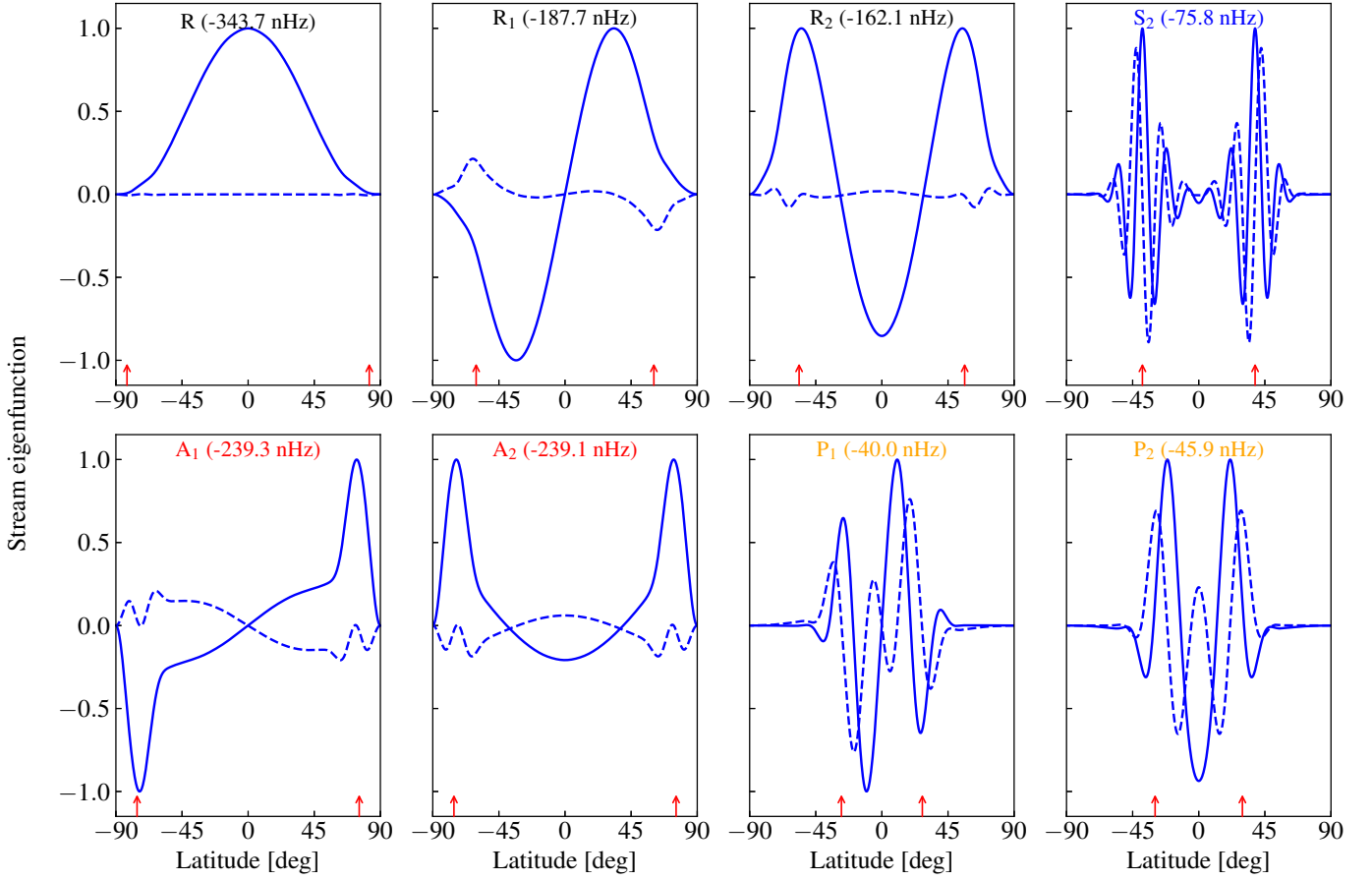}
\caption{Stream functions $\psi$ of selected modes for $m=2$, using the solar surface  differential rotation profile and the Ekman number  $E = 4 \times 10^{-4}$. The solid curves show the real parts, the dashed curves the imaginary parts. In each case, the normalization is such that $\psi=1$ where $|\psi|$ is  maximum (this happens at different latitudes depending on the mode). In each panel, the name of the mode and the real part of its frequency are given (see also Fig.~\ref{fig:spectrum_m1_4}e). The vertical red arrows  give the latitudes of the viscous layers.}
\label{fig:eigenfunctions}
\end{figure*}

\begin{figure*}
\centering
\includegraphics[width=0.9\linewidth]{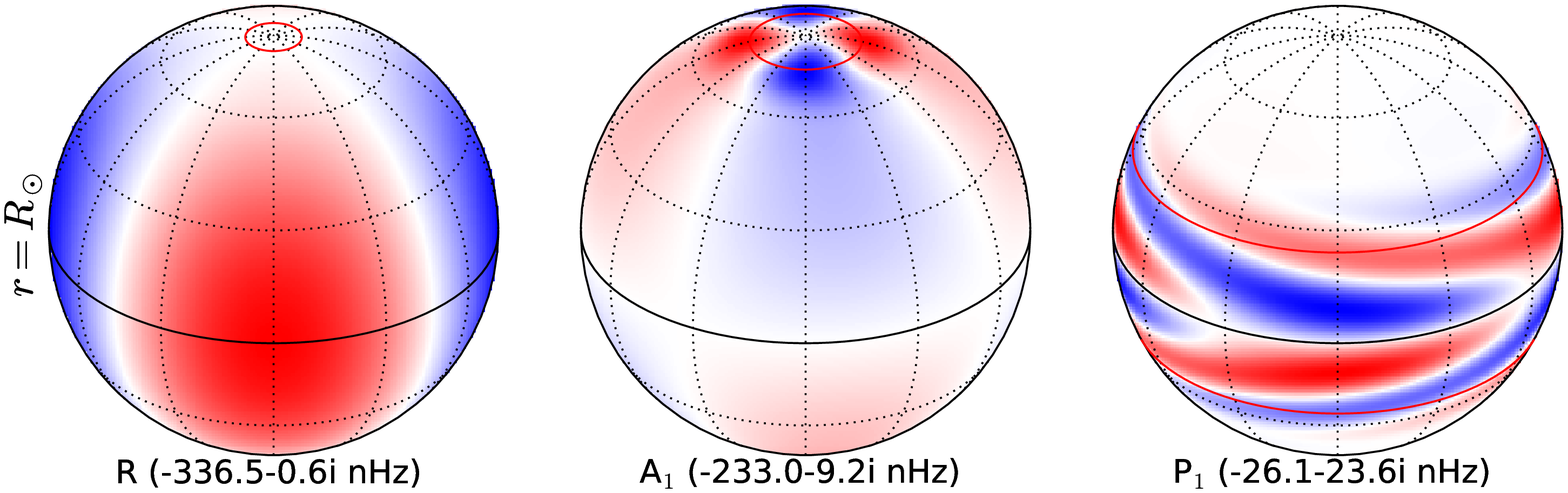} \\
\includegraphics[width=0.9\linewidth]{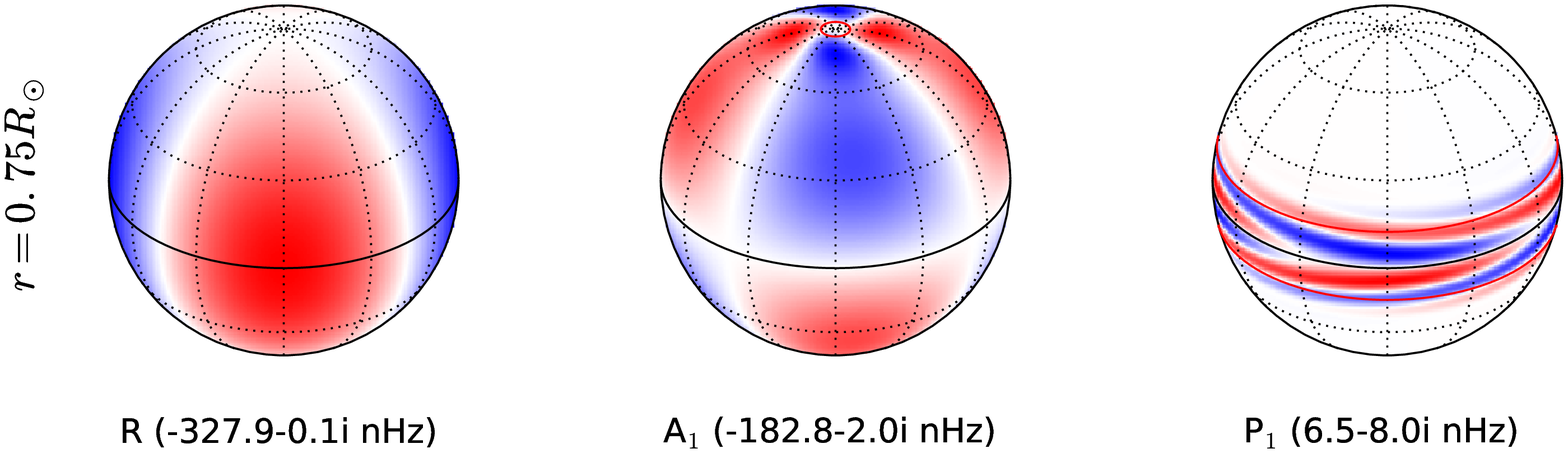}
\caption{Stream functions $\Psi(\theta,\phi,t=0)$ for the $m=2$ modes R, A$_1$ and P$_1$, using the surface solar differential rotation and the Ekman number  $E = 2 \times 10^{-4}$ (top) and at the bottom of the convection zone ($r = 0.75 R_\odot$ with $E = 2 \times 10^{-5}$). Latitudes and longitudes are highlighted every $30^\circ$ with dotted curves and the equator is shown with a solid line. The red curves  show the latitudes of the viscous critical layers. The eigenfrequencies are expressed in the Carrington frame of reference.}
\label{fig:eigenfunctions2D_m2}
\end{figure*}

\begin{figure}
 \centering
\includegraphics[width=0.55\linewidth]{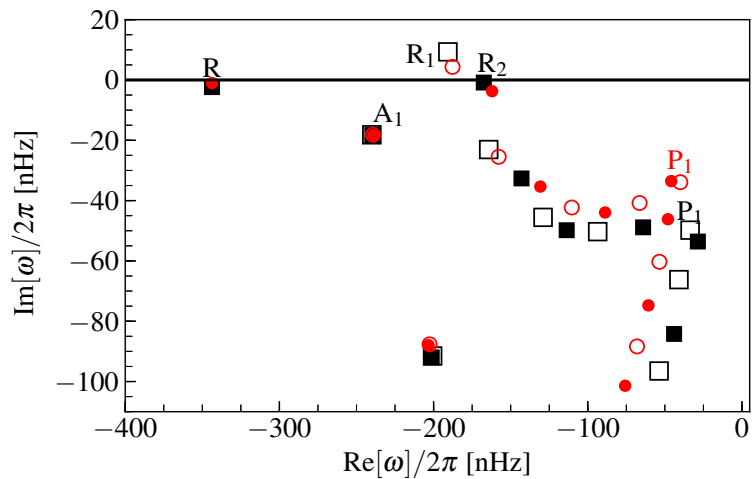}
\caption{Effect of the  meridional flow (Eq.~\ref{Vmer}) on the $m=2$ mode frequencies,  $\omega$. The surface latitudinal  differential rotation is from \citet{Larson2018}.
The black squares show the mode frequencies when  the 
meridional flow is included and the red circles when it is not. 
The full symbols correspond to the symmetric modes and the open symbols to the antisymmetric modes. The Ekman number is fixed at  $E = 4 \times 10^{-4}$ and all frequencies are expressed in the Carrington frame of reference. The eigenfunctions of the modes R, A$_1$, and P$_1$  are shown in Fig.~\ref{fig:eigenfunctions_mer}.}
\label{fig:eigenfrequencies_mer}
\end{figure}

\begin{figure*}
\includegraphics[width=\linewidth]{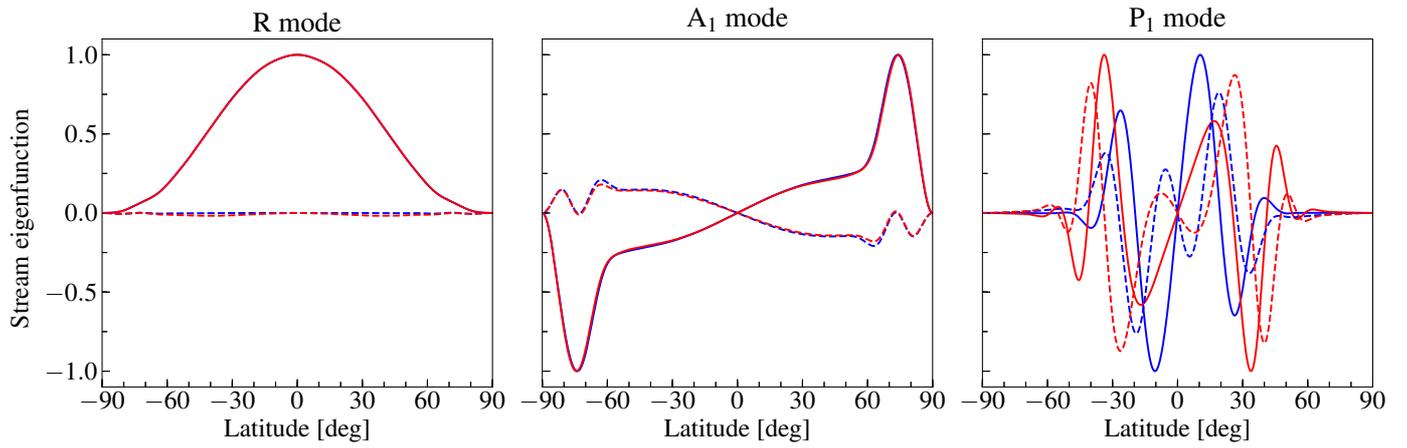}
\caption{Eigenfunctions of the R, A$_1$, and P$_1$ modes with (red) and without (blue)  the meridional flow. The real parts are given by the solid curves and the imaginary parts by the dashed curves. The computations are for $m=2$ and $E = 4 \times 10^{-4}$.}
\label{fig:eigenfunctions_mer}
\end{figure*}

\begin{figure}
    \centering
    \includegraphics[width=0.8\linewidth]{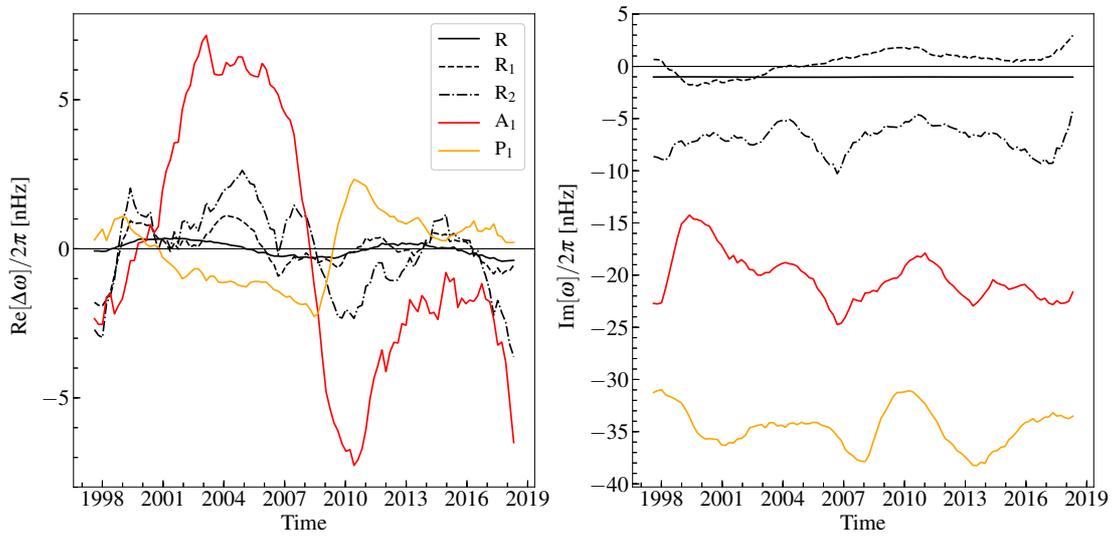} 
    \caption{Temporal changes in  the frequencies $\omega$ of selected $m=2$ modes  caused by the Sun's zonal flows (at the surface). The left panel shows the real parts of the frequencies and the right panel the imaginary parts. The Ekman number is $E = 4 \times 10^{-4}$. Notice that the mode R$_1$ is unstable part of the time.  }
    \label{fig:eigenfrequencyWithTime}
\end{figure}

\begin{figure*}
\centering
\includegraphics[width=0.9\linewidth]{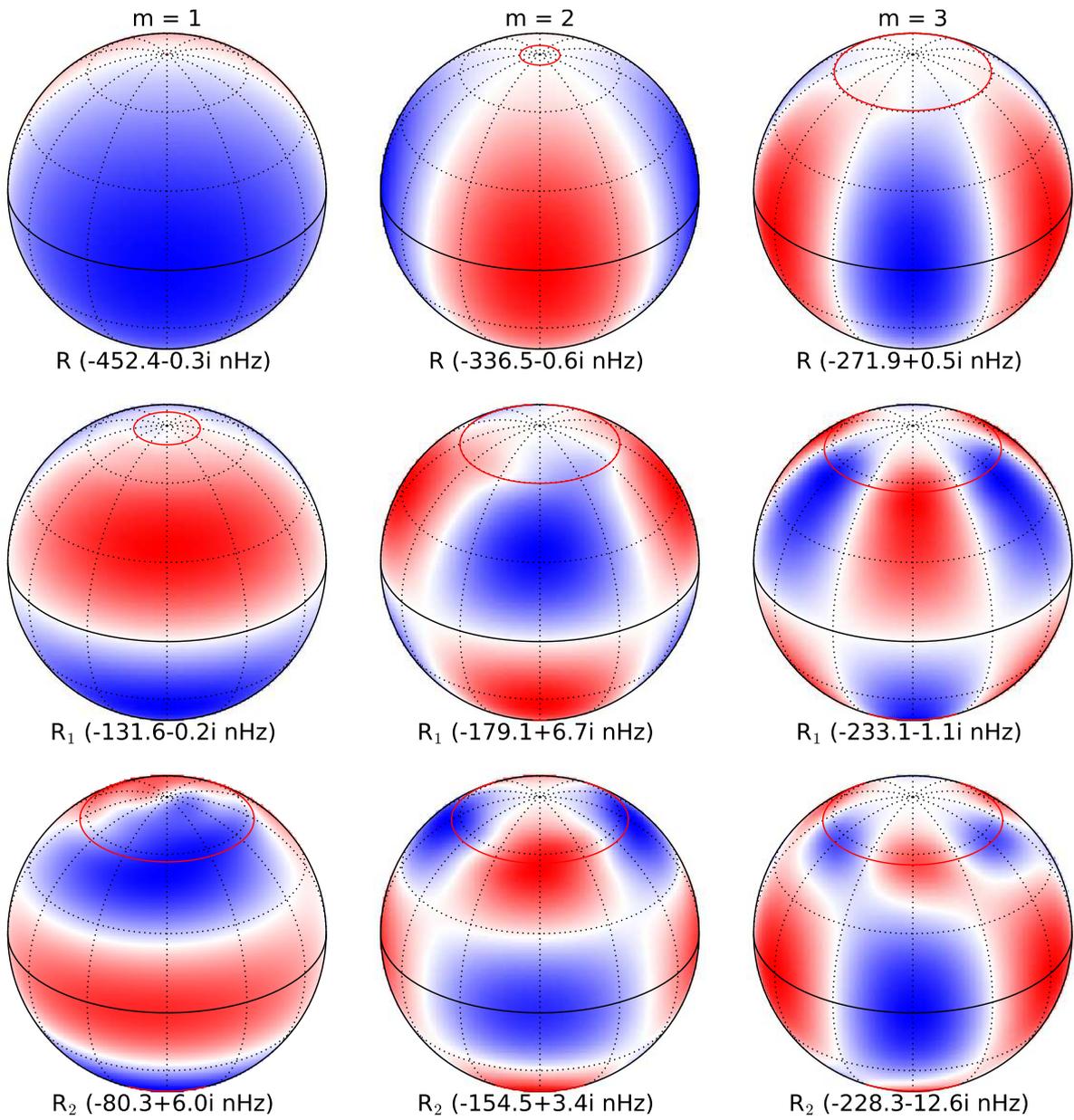}
\caption{Stream functions $\Psi(\theta,\phi,t=0)$ for the $m=$1, 2, and $3$  Rossby modes R, R$_1$, and R$_2$, some of which may be unstable.
The differential rotation is that of the Sun's surface (temporal average) and the Ekman number is $E = 2 \times 10^{-4}$. 
The red curves  show the latitudes of the viscous critical layers. }
\label{fig:eigenfunctions2D_unstable}
\end{figure*}

\begin{figure*}[t]
\centering
\includegraphics[width=\linewidth]{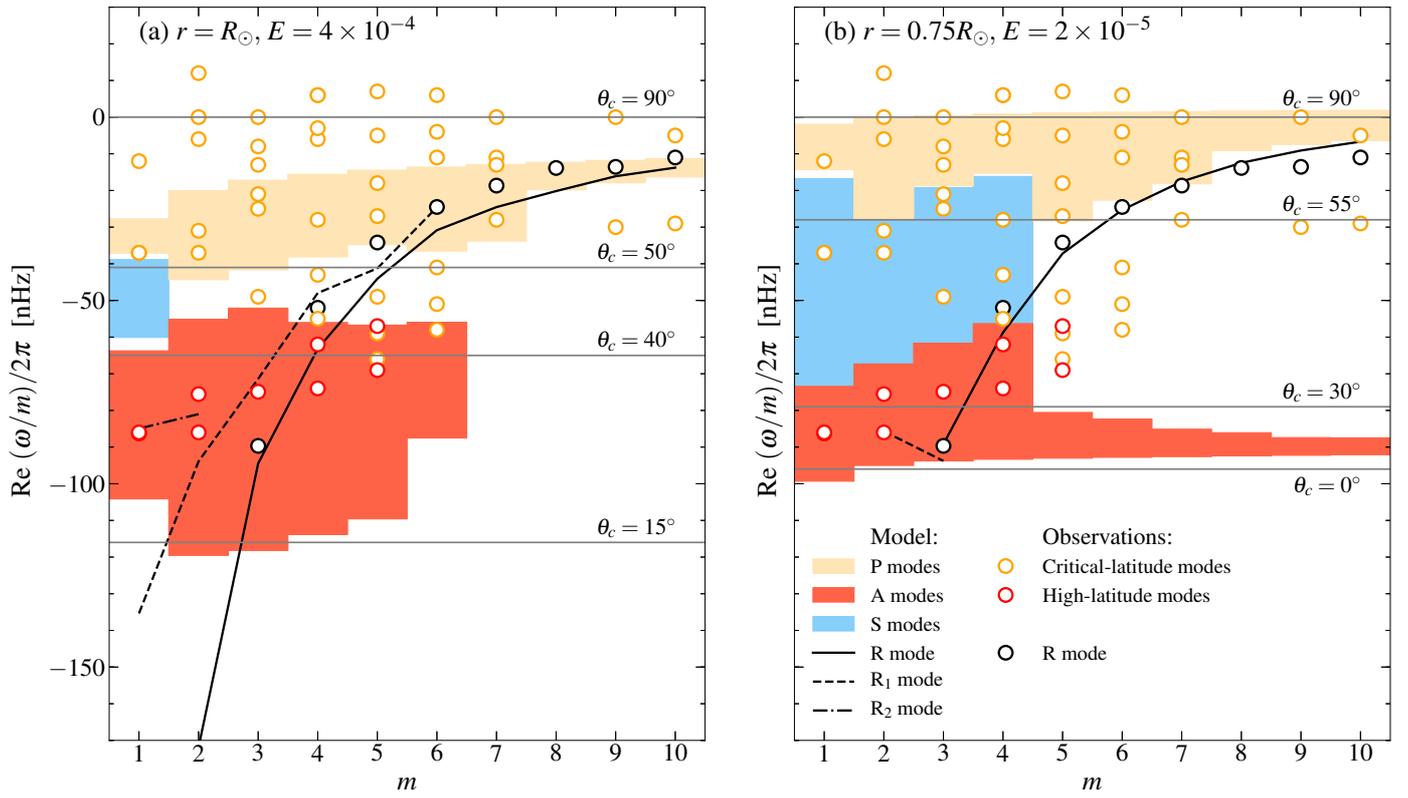} 
\caption{Dispersion relation diagrams for all the modes with  $|\textrm{Im}[\omega]/2\pi|<100$~nHz. This figure is similar to Fig.~\ref{fig:dispersionRelations} but the ordinate  has been replaced by $\omega / m$ so that the critical latitudes occur on horizontal lines. A few selected critical (co-)latitudes $\theta_c$ are highlighted (horizontal grey lines)
}
\label{fig:dispersionRelations_critical}
\end{figure*}

\end{appendix}

\end{document}